\documentclass[preprint]{elsarticle}

\usepackage{amssymb}
\usepackage{amsbsy}
\usepackage{amsmath}
\usepackage{color}
\usepackage{mymacros}
\usepackage{pdflscape}
\usepackage{afterpage}
\usepackage{capt-of}
\usepackage[normal,tight,center]{subfigure}
\usepackage{array}
\usepackage{calc}
\usepackage[thinlines]{easytable}
\usepackage{comment}
\usepackage{tabu}

\def\mysize {\scriptsize \everymath{\displaystyle}}

\journal{Journal of Computational Physics}

\begin{document}

\begin{frontmatter}



\title{High-order asynchrony-tolerant finite difference schemes for partial differential equations}


\author[label1]{Konduri Aditya\corref{cor1}\fnref{fnt1}}
\ead{akonduri@tamu.edu}
\fntext[fnt1]{Current affiliation: Combustion Research Facility, Sandia National Laboratories, Livermore, CA 94550, United States}
\author[label1]{Diego A.\ Donzis}
\ead{donzis@tamu.edu}
\cortext[cor1]{Corresponding author.}
\address[label1]{Department of Aerospace Engineering, Texas A\&M University, College Station, TX 77843, United States}
%\address[label2]{Combustion Research Facility, Sandia National Laboratories, Livermore, CA 94550, United States}

\begin{abstract}
% issue at large scale computing
Synchronizations of processing elements (PEs) in massively parallel simulations, which arise due to communication or load imbalances between PEs, significantly affect the scalability of scientific applications. 
% recent work
We have recently proposed a method based on finite-difference schemes to solve partial differential equations in an asynchronous fashion -- synchronization between PEs is relaxed at a mathematical level.
While standard schemes can maintain their stability in the presence of asynchrony, their accuracy is drastically affected.
% this work
In this work, we present a general methodology to derive asynchrony-tolerant
(AT) finite difference schemes of arbitrary order of accuracy, which can
maintain their accuracy when synchronizations are relaxed. We show that there
are several choices available in selecting a stencil to derive these schemes
and discuss their effect on numerical and computational performance. We provide
a simple classification of schemes based on the stencil and derive schemes that
are representative of different classes.
Their numerical error is rigorously analyzed within a statistical framework to obtain the overall accuracy of the solution. Results from numerical experiments are used to validate the performance of the schemes. 

\end{abstract}


\end{frontmatter}

%\begin{comment}
% \linenumbers

\section{Introduction}
% physics and computations
Numerical simulations are an important tool in understanding complex problems in physics and engineering systems. 
Many of these phenomena are multi-scale in nature, and are governed by nonlinear partial differential equations (PDEs).
With a wide range of scales at realistic conditions, like turbulence phenomena in fluid flows, the numerical solution of these equations becomes computationally very expensive. 
Advances in computing technology have made it possible to carry out intensive simulations on massively parallel computers. 
Currently, state-of-the-art simulations are routinely being done on tens or hundreds of thousands of processing elements (PEs) \cite{DJ2013,DASY2014,MC2016,LMM2013}.

% parallel computing and issues
It is known, at extreme scale, that data communication as well as synchronization between PEs pose a major challenge in the scalability of scientific applications \cite{DBMA+2011}.
In the case of PDE solvers, where the parallelism is typically realized by decomposing the computational domain among PEs, communications that affect the scalability
 arise due to the computation of spatial derivatives in order to propagate the physical 
information across the domain.
The problem becomes more acute in simulations of transient phenomena, where spatial derivatives are evaluated at each time step over an integration of large number of steps.
Another issue concerning the scalability is related to the performance variations across the PEs in a parallel system. 
In this case, sub-optimal performance of even a few PEs may lead to idling of others, as dictated by the data dependencies involved in the computations.
%Note that this issue in turn relates to the communications, as some of the data dependencies are fulfilled by the messages sent between PEs.
It is likely that in future Exascale computing systems, which will have an extremely large PE count, communication and synchronization will be a major bottleneck.
It is thus not surprising that there is a substantial increased interest in developing numerical methods that minimize communications and relax data synchronizations at the mathematical level \cite{GGSZTY2012,Betal2014}. 

% literature and objective
An early effort in solving PDEs in an asynchronous fashion has been presented in \cite{AAIT1994,AAII1996}.
Their method is based on finite-difference schemes and is restricted to the solution of parabolic PDEs with at most second order accuracy.
More recent work \cite{AD2012,DA2014}, again based on finite-difference method, has suggested that due to the randomness in the arrival of messages at different PEs, the resulting algebraic difference equations are stochastic in nature. 
In that work, a statistical framework to analyze such systems was developed to study the numerical properties of commonly used schemes in the presence of asynchrony.
Furthermore, they show that though the stability and consistency of the schemes can be maintained, their accuracy is significantly degraded.
They also proposed the possibility of deriving schemes that are tolerant to communication data asynchrony.
 A follow up of this work to a simple specific equation and numerical scheme has be presented in \cite{Mudigere2014}. 
Although the authors were able to maintain second order accuracy for their chosen scheme when asynchrony is present, one can show using Taylor series that they are severely limited to low order of accuracy.
% higher order schemes will fail to maintain their accuracy.
%At this juncture, it is possible to compute simple problems with lower order accurate asynchronous schemes.
However, as mentioned earlier, a number of natural and engineering systems are multi-scale in nature and will require higher order accurate schemes. 
In this work, we present a general methodology to generate different classes of
high-order asynchrony-tolerant (AT) schemes. This is the main
objective of this paper.

% outline
The rest of the paper is organized as follows. We first briefly review the concept of asynchronous computing for PDEs in section 2. A general method to derive \ats schemes, the choices in stencil available in arriving at these schemes and their classification are presented in section 3. In section 4, we show a statistical framework to analyze the overall accuracy of a numerical method when \ats schemes are used. Numerical experiments to validate the performance of \ats schemes are shown in section 5. Conclusions and further discussions are presented in section 6.

\section{Concept}
\begin{figure}[h]
  \centering
  \includegraphics[width=0.7\textwidth]{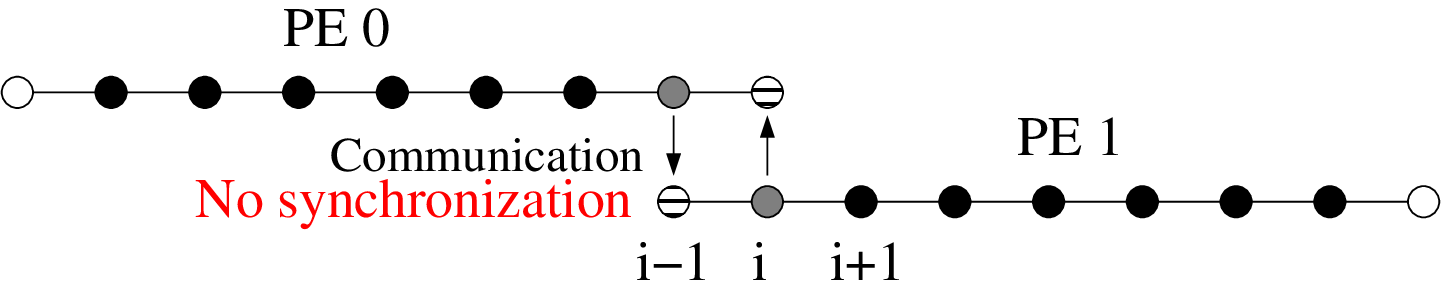}
  \includegraphics[trim=0cm 0cm 0cm 6.6cm,clip=true,width=0.7\textwidth]{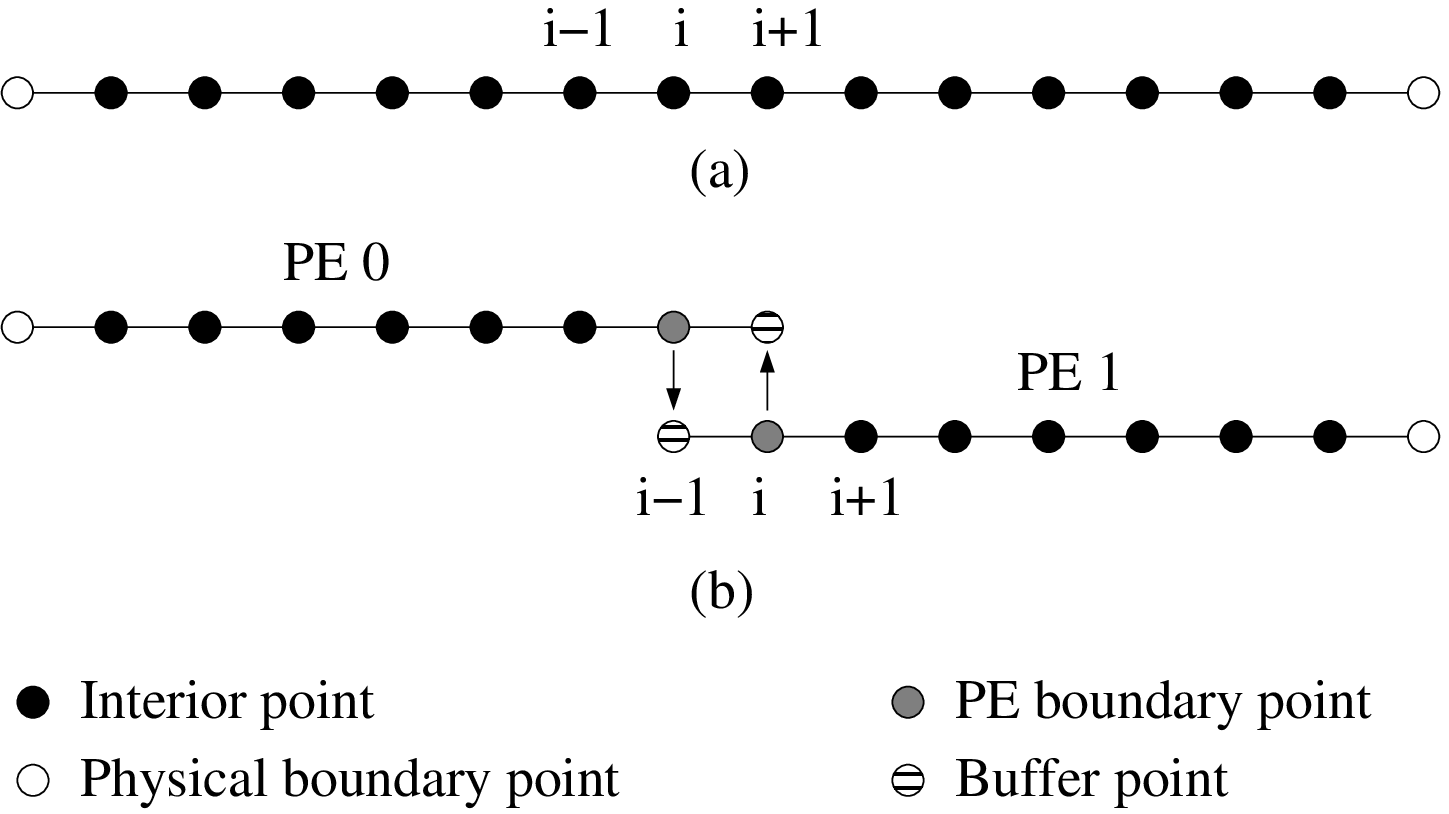}
  \caption{Discretized one-dimensional domain decomposed into two PEs ($P=2$).}
  \label{fig:grid}
\end{figure}

Let $u(x,t)$ be a function of spatial coordinate $x$ and time $t$, which is governed by a time-dependent PDE in a one-dimensional domain.
\rfig{grid} illustrates the discretized domain which is decomposed into $P$ number of PEs. 
Let $i$ and $n$ represent an arbitrary grid point in the domain and time level  such that $u(x_i,t_n)=\U{i}{n}$.
For clarity in the exposition, we assume that the grid points are uniformly distributed in the domain with a spacing $\dx$.
A finite-difference to approximate a spatial derivative at point $i$ and time level $n$ can be expressed, in the most general case, as
\be
\left. \frac{\pd^d u}{\pd x^d} \right|_i^n =
\sum_{j=-\smin}^{\smax} c_{j}
\U{i+j}{n}
 + \ord(\dx^a) ,
\label{eq:general_deriv}
\ee
where $d$ is the order of the derivative,
$\smin$ and $\smax$ are the number of points to the left and right of point $i$ in the stencil, and
$c_j$ is the appropriate coefficient or weight of $\U{i+j}{n}$ such that the scheme is accurate to an order $a$ in space. The term $\ord(\dx^a)$ represents the truncation error of the scheme.

Usually, the numerical solution of a time-dependent PDE is obtained by advancing an initial condition according to an algebraic finite-difference equation in small steps of time $\dt$.
During each time advancement, say, marching from a time level $n$ to $n+1$, spatial derivatives are computed at each grid point using \req{general_deriv}.
% {\colr the values of $u$ at corresponding stencil points}.
In general, these computations are trivial to implement in a serial code, as the value of the function at all the grid points will be locally available in the memory of the PE.
However, if the domain is decomposed into multiple PEs, computations at points near PE boundaries may need values of the function at stencil points that are computed in the neighboring PEs.
Such values are commonly communicated into buffer or 
ghost points,
% {\colr [DD] we need to show in the figure which ones are ``ghost
%points'', no?}, 
as shown in \rfig{grid}.  
Note that the number of values communicated across the left and right PE boundaries is equal to $\smin$ and $\smax$, respectively.

Let $I$ represent the set of physical grid points in the domain and $B$ represent the set of buffer points.
For convenience we divide the set $I$ further such that $I=\Ii\cup\Ib$.
The set of grid points near PE boundaries whose computations need data from neighboring PEs will be denoted by $\Ib$.
The complementary set of interior points, whose computations are independent of communication between PEs is denoted by $\Ii$.
In commonly used parallel algorithms, computations at a point $i\in\Ib$ cannot be advanced until the communication between PEs is complete.
This is typically ensured by enforcing communication synchronization after messages are issued from one PE to another.
As mentioned earlier, with a large number of PEs such synchronizations become expensive and result in poor scalability of codes at extreme scales.
We refer to this as \emph{synchronous} computing.

In the case of \emph{asynchronous} computing, communication between PEs is initiated at each time step, however, the data synchronization is not enforced. 
This means, we cannot ensure that the time level of the function at buffer points is $n$.
It can be $n$, $n-1$, $n-2$, ... depending on the status of messages from successive time advancements.
Due to the random nature of the arrival of messages at different PEs \cite{HSL2008}, the availability of a particular time level at a buffer point is also random. 
Let $\n=n-\k{j}$ be the latest available time level at a buffer point $j$, where $\k{j}$ is the corresponding random delay at that point\footnote{In order to distinguish random
from deterministic variables, we will use a tilde ( $\tilde{}$ ) over the variable for the former.}.
Note that $\n$ can be different at different locations and time levels.
%For simplicity, we assume a uniform delay for all the points that are effected by asynchrony in a stencil.
If we restrict the maximum allowable delay levels to $L$, then $\n\in\left\{n, n-1, ..., n-L+1\right\}$ and $\k{j}\in \left\{ 0, 1 ,...,L-1\right\}$.
The scheme in \req{general_deriv}, when asynchrony is allowed, can be rewritten as
\be
\left. \frac{\pd^d u}{\pd x^d} \right|_i^n \approx
\sum_{j=-\smin}^{\smax} c_{j}
\U{i+j}{n-l}
 ,
\label{eq:general_deriv_async}
\ee
where $l=0$ for $i+j\in I$ and $l=\k{i+j}$ for $i+j\in B$.
Unlike the scheme in \req{general_deriv} which contains a single time level, this scheme uses multiple time levels when some of the points in the stencil belong to the set $B$.
It has been shown in \cite{DA2014} that the accuracy of common
finite-differences used in such an asynchronous fashion is significantly
affected.
In particular, accuracy drops to first order regardless of the original finite
difference used.
Thus, the need to derive AT schemes that maintain accuracy
even when there is a communication delay. We do this next.

\section{Asynchrony-tolerant (AT) schemes}
\subsection{General methodology}
\label{sec:at-gm}
% basic intro and general scheme - 5 
Taylor series and the method of undetermined coefficients provide a systematic procedure to derive finite-difference schemes.
As we show momentarily, this approach can also be used to construct \ats schemes to approximate spatial derivatives.
%When asynchrony is allowed, we do not know, a priori, the latest available time level of the function at buffer points.
%However, the range of values it can take is known.
Let $\U{i+j}{n-l}$ represent the function at a generic point $i+j$ in the stencil with an arbitrary delay of $l$ levels to compute a spatial derivative at a point $i$ and time level $n$. 
Using the $L$ possible time delays, we can express an \ats scheme as
\be
\left. \frac{\pd^d u}{\pd x^d} \right|_i^n \approx
\sum_{j=-\smin}^{\smax}
\sum_{l=0}^{L-1}
\co {j}{l}
%\sum_{j=\smin}^{\smax} \cs{j} \ct{j}{l}
\U{i+j}{n-l}
% + \ord(\dx^a)
,
\label{eq:general_deriv_at}
\ee
where $\co{j}{l}$, for the range of $j$ and $l$, are the appropriate coefficients that have to be determined.
Note that this scheme represents the most general case with the function at all possible time levels at each point in the stencil.
However, depending on the delay at each grid point and time step, which is given by $\k{i+j}$, only one or few time levels may be used in approximating the derivative.
The random nature of $\k{i+j}$ is, now, embedded into $\co{j}{l}$.
The merits of using older time levels not just at buffer points, but also at interior points will be discussed later.

% taylor series - 6
%The coefficients in the scheme expressed in \req{general_deriv_at} can be obtained by solving a system of linear equations that arises by imposing constraints on different terms of the Taylor series, upon expansion of the function at each point and time level in the stencil.
The coefficients in the scheme expressed in \req{general_deriv_at} can be obtained by imposing constraints on different terms of the Taylor series, upon expansion of the function at each combination of point and time level in the stencil.
Let us consider the Taylor series of $\U{i+j}{n-l}$ about the point $i$ and time level $n$.
The series is an expansion in two variables, namely $\dx$ and $\dt$, which is given by
\be
u_{i+j}^{n-l}=\sum_{\pp=0}^\infty \sum_{\qq=0}^\infty u^{(\pp,\qq)} \frac{(j\dx)^\pp (-l\dt)^\qq}{\pp!\qq!},
\label{eq:taylor_series_async}
\ee
where $u^{(\pp,\qq)}$ denotes the $\pp$-th and $\qq$-th partial derivative in
space and time, respectively, of $u$ evaluated at $i$ and $n$.
When $l=0$, the function corresponds to a synchronous value of $u$.
This makes the terms in the series a function of $\dx$ only:
%, as shown below.
\be
u_{i+j}^{n}=\sum_{\pp=0}^\infty u^{(\pp,0)} \frac{(j\dx)^\pp}{\pp!}
\label{eq:taylor_series_sync}
\ee

To obtain the constraints that will assure a given order of accuracy, we substitute the Taylor series of $u$ in the right hand side of \req{general_deriv_at}.
\bea
\sum_{j=-\smin}^{\smax}
\sum_{l=0}^{L-1}
 \co {j}{l}
\U{i+j}{n-l}
&=&
\sum_{j=-\smin}^{\smax} 
\sum_{l=0}^{L-1}
\co {j}{l}
\sum_{\pp=0}^\infty \sum_{\qq=0}^\infty u^{(\pp,\qq)} \frac{(j\dx)^\pp (-l\dt)^\qq}{\pp!\qq!} \nonumber \\
 & = & u^{(0,0)}  \sum_{j=-\smin}^{\smax} \sum_{l=0}^{L-1} \co {j}{l}
  +  u^{(1,0)} \dx \sum_{j=-\smin}^{\smax} \sum_{l=0}^{L-1} j \co {j}{l} - \nonumber \\
  &  &    u^{(0,1)} \dt \sum_{j=-\smin}^{\smax} \sum_{l=0}^{L-1} l \co{j}{l}
-  u^{(1,1)} \dx\dt \sum_{j=-\smin}^{\smax} \sum_{l=0}^{L-1}j l \co {j}{l} + \nonumber \\
& &  \frac{ u^{(2,0)}} {2} \dx^2 \sum_{j=-\smin}^{\smax} \sum_{l=0}^{L-1} j^2 \co {j}{l}
+ \frac{ u^{(0,2)}}{2} \dt^2 \sum_{j=-\smin}^{\smax} \sum_{l=0}^{L-1} l^2 \co{j}{l}
+ \dots
\label{eq:exp}
\eea
The linear combination of the function values, in the above equation, represents a scheme when ($i$) the coefficient of the $d$-th derivative of $u$ in space is unity, and ($ii$) low-order terms are eliminated according to the desired accuracy of the scheme.

Let $a$ be the desired order of accuracy in space.
This means that the leading order term in the truncation error should vary with the grid spacing as $\dx^a$.
In the Taylor series of the function for synchronous schemes, as in \req{taylor_series_sync}, the lower order terms can be readily identified as the ones with the power of $\dx$ less than $d+a$.
However, when asynchrony is present this is not obvious. The terms in the series can now be a function of either or both $\dx$ and $\dt$, which are usually not independent.
In order to identify the lower order terms, let us assume the relation $\dt\sim\dx^r$.
Such a relation is often obtained from analysis of the scheme's numerical stability or other constraints posed by the physics of the problem.
Using this relation, we can arrive at the condition to identify lower order terms that need to be eliminated to obtain a scheme of order $a$. This expression is: $\pp+r\qq<d+a$.
% conditions - 8
Using \req{exp}, we can then summarize the constraints as
\be
 \sum_{j=-\smin}^{\smax} \sum_{l=0}^{L-1} \co{j}{l} \frac{(j\dx)^\pp (-l\dt)^\qq}{\pp!\qq!} =
  \begin{cases}
    1       & \quad \text{for } (\pp,\qq) = (d,0) \\
    0       & \quad \text{for } \pp+r\qq<d+a; (\pp,\qq) \ne (d,0). \\
  \end{cases}
\label{eq:cond}
\ee
Clearly, the first condition in the above equation makes the coefficient of the term corresponding to $d$-th spatial derivative on the right hand side of \req{exp} unity.
The second condition will set to zero all the necessary lower order terms to obtain an overall accuracy $a$.
For a given stencil, these conditions give rise to a system of linear equations. 
The number of equations in the system is one more than the number of lower order terms that have to be eliminated from \req{exp}.
Let $\bs{A}\bs{\tilde{c}}=\bs{b}$ represent this system, where $\bs{A}$ is the coefficient matrix whose elements are a function of $j$ and $l$, $\bs{\tilde{c}}$ is the vector of variables that contains coefficients in the scheme and $\bs{b}$ is the vector with zero elements except for the row corresponding to the order of the derivative to be approximated.
The solution to this system determines the coefficients of the scheme.
% SMALL SIZE LS
%In the method outline above, conditions resulting from \req{cond}, for a given stencil, are all solved in a single linear system.
%However, this is not essential and one can satisfy these conditions in several stages. In these cases, we can create select few terms and solve in different linear systems.
%
%In general, it is useful to have asynchrony-tolerant schemes that are close to or that resemble commonly used schemes under synchronous conditions. Such schemes, as we show in example 3, need a specific stencil pattern and conditions on the coefficients. 
%To implement these considerations in deriving a scheme,  

\begin{figure}[h]
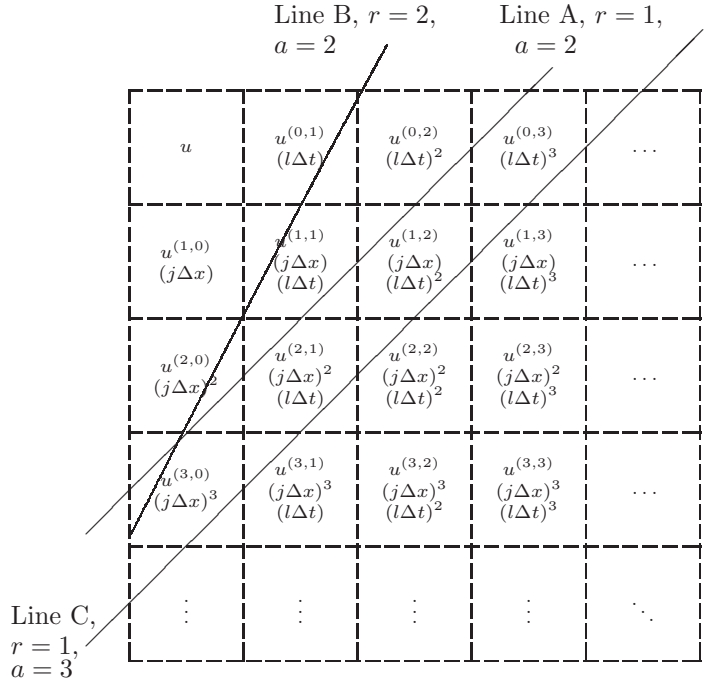

\begin{center}
{\ss
\begin{TAB}(e,1.5cm,1.5cm){:c:c:c:c:c:}{:c:c:c:c:c:}
\bmpc${u}$\empc    &     {\bmpc ${u^{(0,1)}}$ ${ \ldt}$\empc}     &     {\bmpc$u^{(0,2)}$ $\ldt^2$\empc}     &     {\bmpc$u^{(0,3)}$ $ \ldt^3$\empc}    &    { $\dots$} \\
{\bmpc${u}^{(1,0)}$ $\ldx$\empc }     &     {\bmpc${u^{(1,1)}}$  ${\ldx}$\\ ${\ldt}$ \empc}      &    {\bmpc$u^{(1,2)}$  $\ldx$ \\ $\ldt^2$\empc}      &     {\bmpc$u^{(1,3)}$  $\ldx$ \\ $\ldt^3$\empc}      &     { $\dots$} \\
{\bmpc${u}^{(2,0)}$ $\ldx^2$\empc }     &     {\bmpc${u^{(2,1)}}$  ${\ldx^2}$\\ ${\ldt}$ \empc}      &    {\bmpc$u^{(2,2)}$  $\ldx^2$ \\ $\ldt^2$\empc}      &     {\bmpc$u^{(2,3)}$  $\ldx^2$ \\ $\ldt^3$\empc}      &     { $\dots$} \\
{\bmpc${u}^{(3,0)}$ $\ldx^3$\empc }     &     {\bmpc${u^{(3,1)}}$  ${\ldx^3}$\\ ${\ldt}$ \empc}      &    {\bmpc$u^{(3,2)}$  $\ldx^3$ \\ $\ldt^2$\empc}      &     {\bmpc$u^{(3,3)}$  $\ldx^3$ \\ $\ldt^3$\empc}      &     { $\dots$} \\
{$\vdots$ } & {$\vdots$ } & {$\vdots$ } & {$\vdots$ } & {$\ddots$ } \\
\end{TAB}
}
{\setlength{\unitlength}{1cm}
\bp
% r=1
\put(-8.5,1.7){\line(1,1){6.2}}
\put(-3,8.5){Line A, $r=1$,}
\put(-2.8,8.1){$a=2$}
% r=1
\put(-8.5,0.2){\line(1,1){8.2}}
\put(-9.5,0.5){Line C,}
\put(-9.5,0.1){$r=1$,}
\put(-9.5,-0.2){$a=3$}
%r=2
\qbezier(-7.9,1.7)(-5.7,6.03)(-4.5,8.2)
\put(-6.0,8.5){Line B, $r=2$,}
\put(-6.0,8.1){$a=2$}
\ep}
\caption{Terms in the Taylor series of $\U{i+j}{n-l}$ illustrated in a matrix format. Constant in each term are omitted for clarity. Lines A, B and C represent $\pp+r\qq=d+a$ for different sets of parameters.}
\label{fig:ts1}
\end{center}
\end{figure}

% linear system and asynchrony - 9
Before getting into the discussion on the choice of stencil, we make a few observations regarding the linear system when asynchrony is present.
To aid the discussion we express the terms in the Taylor series of the function at the generic stencil point, $\U{i+j}{n-l}$, in a matrix format as shown in \rfig{ts1}.
This provides a simple format to visualize different terms in the series and help us easily identify the terms on which the conditions in \req{cond} have to be imposed.
%\vspace{1cm}
In this graphical representation, we omit constants in each term for the sake of clarity.
%We can now easily identify the terms affected by \req{cond}.
%We know that the conditions in \req{cond}, imposed on the terms from Taylor series, result in a set of linear equations.

%The conditions are imposed on
In words, \req{cond} implies constraints on the term containing the derivative of order $(d,0)$ and on all the terms that satisfy the inequality $\pp+r\qq<d+a$, that is, all terms above the $\pp+r\qq=d+a$ line in \rfig{ts1}.
With this representation, we can easily separate terms that need to be eliminated from those that do not.
To illustrate this, let us choose $d=1$, $a=2$ and $r=1$, which corresponds to a second-order approximation of the first derivative, using a convective-type CFL condition such that $\dt\sim\dx$.
For these parameters, conditions are imposed on the terms with $(\pp,\qq)=\{(0,0),(1,0),(2,0),(0,1),(1,1),(0,2))\}$, which are the terms above the line $A$ in the figure.
If asynchrony is absent, that is, $l=0$, the only terms that are non-zero in the table belong to the first column. 
%The conditions, now, have to be imposed only on the non-zero terms above the line $A$.
This shows that, for a given accuracy, the number of terms on which 
conditions are imposed is larger when asynchrony is present, which thus results in a larger linear system.

The increase in the number of equations also depends on $r$, which relates $\dt$ and $\dx$.
For example, the situation for $r=2$ is also shown in \rfig{ts1} with line $B$.
The number of terms above the line $B$ (4 terms) is less than $A$ (6 terms),
which implies that a higher $r$ will reduce the number of lower order terms due
to asynchrony for a given accuracy.

The other aspect is the increase in stencil size with increase in accuracy.
In commonly used synchronous schemes ($l=0$), a successive increase in the
order of accuracy will impose a new condition on one more term in the Taylor
series,
%{\colb
which adds an additional equation to the linear system. The linear system is then solved by adding one more grid point to the stencil.
%}
%{\colb
%The situation of adding a grid point to the case of line $A$
%is shown as line $C$ in \rfig{ts1}.
%This adds an additional equation to the linear system.
%} 
% {\colr [DD]: I changed this paragraph but I'm not sure what we are 
% trying to say here. What is really line C?}
%that can be solved by
%adding one more grid point to the stencil.
%For example, when we increase the accuracy of a scheme, $a$, from $2$ to $3$ for $d=1$, the term with $(\pp,\qq)=$ 
However, in deriving \ats schemes more than one additional equation may be
added to the system (compare the number of terms above lines A and C). 
Thus, we expect the stencil of \ats schemes to grow larger than commonly used
synchronous schemes when the desired accuracy is increased.

\subsection{Choice of stencil}
\label{sec:choice}
% stencil - 10
In principle one can choose a stencil that consists of different grid points
and time levels to approximate spatial derivatives.
However, the stencil of commonly used synchronous schemes are constructed
exclusively with spatial grid points.
This has some advantages.
First, the function at the synchronous time level is available for spatial
derivative evaluation at all points in the domain.
Second, as argued in \cite{DA2014}, and elaborated in \rsec{at-gm} above, this
choice avoids the additional terms that will appear in Taylor series when the
stencil consists of delayed time levels.

When asynchrony is present, on the other hand, as is clear from \req{general_deriv_at}, the function can belong to multiple time levels.
Thus, we can take advantage of using the function at delayed time levels in deriving \ats schemes.
%The schemes can have the function at different grid points, as well as, different time levels at each grid point.
The choice of stencil should be made according to the nature of terms in Taylor series on which conditions in \req{cond} are imposed.
To understand this let us recall the tabular representation of the Taylor series of $\U{i+j}{n-l}$, as shown in \rfig{ts2}.
\begin{figure}[h]
\begin{center}
{\ss
\begin{TAB}(e,1cm,1cm){:c:c:c:c:c:}{:c:c:c:c:c:}
\bmpc${u}$\empc    &     {\cg\bmpc ${u^{(0,1)}}$ ${ \ldt}$\empc}     &     {\cg\bmpc$u^{(0,2)}$ $\ldt^2$\empc}     &     {\cg\bmpc$u^{(0,3)}$ $ \ldt^3$\empc}    &    {\cg $\dots$} \\
{\cb \bmpc${u}^{(1,0)}$ $\ldx$\empc }     &     {\colr \bmpc${u^{(1,1)}}$  ${\ldx}$\\ ${\ldt}$ \empc}      &    {\colr \bmpc$u^{(1,2)}$  $\ldx$ \\ $\ldt^2$\empc}      &     {\colr \bmpc$u^{(1,3)}$  $\ldx$ \\ $\ldt^3$\empc}      &     {\colr $\dots$} \\
{\cb $\vdots$ } & {\colr $\vdots$ } & {\colr $\vdots$ } & {\colr $\vdots$ } & {\colr $\ddots$ } \\
{\cb \bmpc ${u}^{(d,0)}$ $\ldx^d$ \empc}      &    {\colr \bmpc ${u^{(d,1)}}$ $\ldx^d$ \\ $\ldt$ \empc}      &     {\colr \bmpc $u^{(d,2)}$ $\ldx^d$ \\ $\ldt^2$ \empc}      &     {\colr \bmpc $u^{(d,3)}$ $\ldx^d$ \\ $\ldt^3$ \empc}      &   {\colr $\dots$} \\
{\cb $\vdots$ } & {\colr $\vdots$ } & {\colr $\vdots$ } & {\colr $\vdots$ } & {\colr $\ddots$ } \\
\end{TAB}
}
\caption{Terms in the Taylor series of $\U{i+j}{n-l}$ illustrated in a matrix format. Constants in each term are omitted for clarity. Different colors represent terms from different groups, as explained in \rsec{choice}.}
\label{fig:ts2}
\end{center}
\end{figure}
We can classify terms into four groups, as represented by the four different colors in the figure.
Terms in blue are a function of $\dx$ alone. These terms will appear in the Taylor series of the function when $j\ne 0$.
Similarly, terms that are a function of $\dt$ only are shown in green and they appear when $l\ne 0$.
Terms in red are a function of both $\dx$ and $\dt$, and these appear when $j\ne0$ and $l\ne 0$.
The term $u$ in black is a function of neither $\dx$ nor $\dt$ and is present in the Taylor series of the function at any point and time level.
In order to eliminate specific terms in the truncation error, it is apparent that we cannot arbitrarily choose the points and time levels in a stencil.
They have to be selected according to the number of terms in each of these groups.
For example, if a linear system consists of three equations that correspond to condition on terms that belong to the red group, then the scheme would need the function evaluated at a minimum of three combinations of $j$ and $l$ such that $j\ne0$ and $l\ne 0$. 
If not, the linear system may not have a solution or may have a solution which correspond to stencils completely biased towards the synchronous side of the stencil, like forward and backward differences.
%, we would not be able to solve the linear system or may end up with a scheme that uses one-sided stencil with synchronous time level like forward or backward differences.

% computational performance
The choice of stencil has consequences also in terms of the performance of simulation codes on parallel machines. 
%With respect to performance metrics on parallel machines, 
Expanding the stencil in space will lead to larger message sizes to be sent over the network, which may be too expensive at extreme scales.
Using multiple levels in time will keep the messages relatively smaller, but will increase the memory requirements in each PE.
This choice, thus, would require information on the specific computing system to be used for the simulation.
% gives us the choice to expand the stencil according to computing machine characteristics.

\begin{figure}[h]
  \centering
  \includegraphics[width=0.5\textwidth]{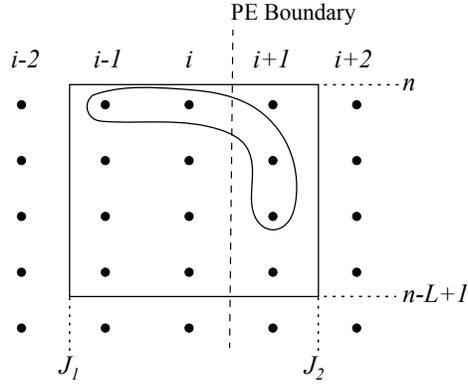}
  \caption{A schematic of stencil layout for a particular
  asynchrony-tolerant (AT) scheme.}
  \label{fig:layout}
\end{figure}
The rectangular box in \rfig{layout} illustrates the layout of the stencil used in expressing the general scheme in \req{general_deriv_at}.
However, as mentioned earlier, not all the time levels at all points are required to approximate the derivative.
Instead, one can limit the number of time levels at each grid point in such a way to introduce the exact number of coefficients that would make the linear system solvable. \req{general_deriv_at}, then, becomes:
%With the number of equations known in the linear system from previous section, we can determine the exact size of the stencil in terms of the number of points and the number of time levels at each point. 
%In the case of buffer points, as we do not know, a priori, the latest available time level, we can proceed in deriving the scheme using the delay $\k{i+j}$.
%With a specific choice of stencil from the layout, the general scheme in \req{general_deriv_at} can be rewritten as
\be
\left. \frac{\pd^d u}{\pd x^d} \right|_i^n \approx
\sum_{j=-\smin}^{\smax}
\sum_{l=\lmin{j}}^{\lmax{j}}
\co{j}{l}
\U{i+j}{n-l}
% + \ord(\dx^a)
,
\label{eq:general_deriv_at2}
\ee
where \lmin{j}\ and \lmax{j}\ are the lower and upper limits on the time levels used at the point $i+j$.
These limits are computed according the latest time level available and the number of time levels chosen at that point in the stencil.
As an example, in \rfig{layout} we identify a stencil to solve a system with four equations. At the two interior points the latest available time level is $n$, which has a zero delay.
Thus, the limits are $\lmin{j}=\lmax{j}=0$, for $j\in \{-1,0\}$.
As specified before, the latest available time level at the buffer point is given by $\n=n-\k{i+1}$, and we use two successive time levels at this point. The limits on the time level at this point are then $\lmin{1}=\k{i+1}$ and $\lmax{1}=\k{i+1}+1$.  

% numerical aspects
A choice of stencil will lead to a scheme only when there exists a solution to the resulting linear system $\bs{A}\bs{\tilde{c}}=\bs{b}$.
Since $\bs{b}\neq \bs{0}$ due to the first condition in \req{cond}, the system is non-homogeneous and has a unique solution only when the matrix $\bs{A}$ is non-singular or has a full rank.
If $N_A$ is the size of the linear system, then the matrix has full rank when $rank(\bs{A})=N_A$.
We, then, obtain the scheme by solving the system and substituting the coefficients into \req{general_deriv_at2}.
On the other hand, when $rank(\bs{A})<N_A$, the matrix is singular and  
the linear system possesses either no solution or infinite solutions.
We can distinguish these two cases by computing the rank of the augmented matrix $\bs{A}|\bs{b}$.
If $rank(\bs{A})\ne rank(\bs{A}|\bs{b})$, then the system is inconsistent and the choice of stencil does not result in a scheme.
In the case where $rank(\bs{A})= rank(\bs{A}|\bs{b})$, the linear system is consistent, but has infinite solutions.
The linear system, then, contains two or more equations that are linearly dependent.
This means that for the choice of stencil, conditions on at least two of the terms are mathematically equivalent or a condition on at least one of the terms can be obtained from linear combination of others.
In such a situation, we can get a scheme with greater accuracy with the same stencil.
%This is achieved by imposing conditions on the next few high order terms in Taylor
In some cases, it is possible to construct a smaller linear system (with a corresponding smaller stencil) comprised of linearly independent equations, which does in fact have a unique solution.
The greater the number of linearly dependent equations, the smaller the linear system with linearly independent equations will be.
This suggests that a judicious selection of grid points and time levels can be used to increase the number of linearly dependent equations in the resulting system, and thus reduce the stencil size which in turn reduces computations as well as the size of communication messages.
%For these reasons, identifying stencils that will lead to linearly dependent equations will be of interest in deriving asynchrony-tolerant
This will be of interest in deriving \ats schemes, which demand larger stencil due the presence of terms due to asynchrony.

Let us recall the second condition from \req{cond}, imposed to eliminate the terms due to asynchrony in deriving a scheme. After cancelling out the term $\dx^\pp \dt^\qq / \pp! \qq! $, which is constant across the equation for a given $(\pp,\qq)$, we get
\be
 \sum_{j=-\smin}^{\smax} \sum_{l=\lmin{j}}^{\lmax{j}} \co{j}{l} j^\pp l^\qq = 0.
\label{eq:cond-async1}
\ee
It is evident from the above equation that the existence of linearly dependent equations rests on the values of $j$ and $l$ which are defined by the stencil, as well as $\pp$ and $\qq$ which represent the order of the derivative corresponding to the equation.
As mentioned earlier, the function in the stencil can belong to multiple time levels.
The time level of the function at the interior points has a zero delay, that is $l=0$, and hence, will not appear in equations corresponding to asynchrony terms.
If we choose a single uniform time level with a delay $\k{i+j}=\tilde{k}$ for all $i+j \in B$ in the stencil, then $\lmin{j}=\lmax{j}=\tilde{k}$ which leads to a uniform value of $l^\qq$ in \req{cond-async1}. 
The equation then reduces to
\be
 \sum_{i+j\in B} \co{j}{\tilde{k}} j^\pp = 0,
\label{eq:cond-async2}
\ee
which is independent of $\qq$, and shows that for a given $\pp$, equations corresponding to $\qq \ne 0$ are linearly dependent.
With reference to \rfig{ts2}, when we eliminate a term in the red or green groups, all the other terms in the corresponding row that are in the same group are also eliminated.
This illustration shows that it is possible to choose a stencil which results in linearly dependent equations in a system.
%Of course, there are other ways to obtain such systems, but we will not discuss them here.

% guidelines
We conclude this section by summarizing the steps to derive \ats schemes.
\begin{enumerate}
\item List the terms on which conditions have to be imposed for a given $d$, $a$ and $r$
\item Identify an appropriate stencil ($\smin,\smax,\lmin{j},\lmax{j}$) according to the terms in the list
\item Compute the rank of the matrix $\bs{A}$
\bi
\item $rank(\bs{A})=N_A$: unique solution
\item $rank(\bs{A})<N_A$ and $rank(\bs{A})\ne rank(\bs{A}|\bs{b})$: no solution, identify a new stencil
\item $rank(\bs{A})<N_A$ and $rank(\bs{A})= rank(\bs{A}|\bs{b})$: infinite solutions, add more conditions to get greater accuracy or reduce the system size with only linearly independent equations and adjust the stencil size
\ei
\item  Solve for $\bs{\tilde{c}}$ and substitute the coefficients into general scheme
\end{enumerate}

\subsection{Alternative approach} \label{sec:alt-app}
It is often necessary to use schemes with a specific structure in terms of stencil and the corresponding coefficients to either improve computational performance or satisfy numerical properties.
In the context of \ats schemes, it is desirable to use schemes at PE boundary points that are similar in nature to those at interior points.
Such an implementation may improve the overall stability of a numerical method and relieve the natural tendency of concentrated errors in the spatial distribution near PE boundaries (e.g. see Fig. (3) in \cite{DA2014}).
Though the method described earlier gives the flexibility to choose a particular structure for the stencil, there is not much control over the nature of the resulting coefficients in the scheme.
This is because the necessary conditions imposed on the terms on the right hand side of \req{exp} are all solved in a single linear system.
And, explicit conditions on the coefficients have to be added to the linear system to address this issue.

In an alternative approach to derive schemes, we propose to impose necessary conditions (similar to \req{cond}) on the set of terms arising from Taylor series in a step-by-step process.
In each step, a subset of lower order terms are eliminated, while retaining the derivative order term using a particular stencil.  
This process is repeated until the desired accuracy is achieved.
Linear systems of smaller size can be constructed in each step to enforces the conditions and obtain the coefficients.
The procedure described in bullet $3$ in the summary of \rsec{at-gm} should be used in computing the solution of these systems.

We now proceed to outline the procedure to derive schemes similar to central differences using this approach, and will later provide a detailed illustration in example 3 of \rsec{class}.

Central difference schemes are widely used in solving parabolic and elliptic PDEs, and are shown to have low numerical dissipation, necessary to resolve all scales in multi-scale phenomena \cite{hirsch.I}.
If we consider the structure of central difference schemes, it can be
characterized by a symmetric stencil about the point of computation, and
symmetric coefficients (in absolute value).
A general synchronous central difference scheme can be expressed as 
\be
\left. \frac{\pd^d u}{\pd x^d} \right|_i^n \approx
\sum_{j=0}^{J}
\phi_{j}
\left( \U{i+j}{n} + (-1)^d \U{i-j}{n} \right)
% + \ord(\dx^a)
,
\label{eq:general_deriv_cds}
\ee
where $J$ determines the size of stencil and $\phi_j$ are the appropriate coefficients.
Let us consider this stencil in the presence of asynchrony. 
In practical simulations each PE, typically, is assigned a large number of grid points.
% This is due to the fact that computation rates are faster than communication \cite{DBMA+2011}.
When asynchrony is allowed in such cases, delays are experienced only on one side of the stencil, that is, either on the left or right about the point of computation $i$ for $i\in I_B$.
If we assume the delay on the left side, which implies $i-j$ is a buffer point, then the terms in the sum in \req{general_deriv_cds} take the form $\left( \U{i+j}{n}+ (-1)^d\U{i-j}{n-l} \right)$.
To maintain the above mentioned symmetries in \ats schemes, we use this sum to eliminate some of the lower order terms and retain the derivative order term in the Taylor series in the first step.
%Note that to make a consistent comparison
%We consider coefficients to be symmetric, if they are so when $l=0$, that is, when there is a zero delay.
As delay is present only at $i-j$, none of the terms due to asynchrony in the expansion of $\U{i-j}{n-l}$ are cancelled out in the sum of the function at the two points.
However, some of the terms, which are not a function of $\dt$, cancel out depending on the order of derivative $d$.
If $d$ is odd, then terms that correspond to even power of $\dx$ cancel out, as shown next.

Consider the difference $\U{i+j}{n}- \U{i-j}{n}$ where we choose $l=0$ to simplify the analysis.
The conclusions, though, are valid for arbitrary delays $l>0$. A Taylor series expansion can then be written as
\be
\U{i+j}{n}- \U{i-j}{n} = 2 \left[ u^{(1,0)} \frac{(j\dx)}{1!} + u^{(3,0)} \frac{(j\dx)^3}{3!} + u^{(5,0)} \frac{(j\dx)^5}{5!} + \dots \right]
\label{eq:deriv11}
\ee
Similarly, if we consider the sum, $\U{i+j}{n}+\U{i-j}{n-l}$, terms with odd powers of $\dx$ will vanish.
This reduces some of the terms on which conditions need to be imposed, as we move on to the next step.
A further decrease in the number of conditions can be achieved by artificially imposing the same delay of $l$ levels on the other side of the stencil, that is, $\left( \U{i+j}{n-l} + (-1)^d \U{i-j}{n-l} \right)$.
The Taylor series expansion of this difference, for odd $d$, is
\bea
\U{i+j}{n-l}-\U{i-j}{n-l} & = & 2 \left[ u^{(1,0)} \frac{(j\dx)}{1!} + u^{(1,1)} \frac{(j\dx)(-l\dt)}{1!1!} \right. \nonumber\\
& & \left. + u^{(3,0)} \frac{(j\dx)^3}{3!} + u^{(1,2)} \frac{(j\dx)(-l\dt)^2}{1!2!}  + \dots \right],
\label{eq:deriv22}
\eea
which shows that all the terms with even powers of $\dx$, regardless of the power of $\dt$, are absent.
Indeed, imposing this artificial delay on the function at the interior point, though demands additional storage of more time levels at each grid point, will lead to a smaller number of constraints. Thus, schemes with delay on both sides will need a smaller stencil to compute derivatives.

In \cite{Mudigere2014}, this approach was used to recover the drop in accuracy due to delay in communication in a particular application using central difference schemes.
The authors further suggested that imposing delay on both sides of the stencil in central differences would suffice to maintain the accuracy under asynchronous conditions.
However, it can be shown from a Taylor series expansion that the schemes cannot be accurate beyond second order under the conditions they presented.
It is essential to increase the stencil size to achieve higher order accuracy when asynchrony is present, as shown in this work.
%Thus, we move onto the next step to eliminate the remaining higher order terms.

The remaining lower order terms can be eliminated by expanding the stencil with additional terms of the form $\left( \U{i+j}{n}+ (-1)^d\U{i-j}{n-l} \right)$ for different values of $j$ or $l$.
This can be done either in a single or multiple steps, and both of them will ensure symmetry in the coefficients.
Assuming delay on the left of the stencil, the resultant \ats scheme takes the
form: 
\be
\left. \frac{\pd^d u}{\pd x^d} \right|_i^n \approx
\sum_{j=0}^{J}
\sum_{l=\lmin{-j}}^{\lmax{-j}}
\cats{j}{l}
\left( \U{i+j}{n} + (-1)^d \U{i-j}{n-l} \right)
% + \ord(\dx^a)
\label{eq:general_deriv_cds-at}
\ee

It is often useful to derive \ats schemes that reduce to central difference schemes when all delays are zero, that is $\k{j}=0$ for $j\in B$.
%Let us consider the remaining higher order terms that have to be eliminated to obtain the scheme.
%These terms will be a function of $\dx$, $\dt$ and $\dx$ and $\dt$.
Such schemes can be derived by expanding the stencil in space, using the sum at different $j$, to eliminate terms that are not a function of $\dt$. And, use the sum at different levels in time to cancel out the terms due to asynchrony.
%, then the resulting scheme will resemble central differences.
%We illustrate this procedure in detail in Example 3.
This approach, which contains the essence of the alternative procedure
presented in this section, will be illustrated in detail as Example 3 below.

\subsection{Classification of schemes}\label{sec:class}
% intro - 5
In arriving at \ats schemes there are several choices available in terms of choosing points and time levels in a stencil and on the nature of coefficients. We first provide a simple classification of \ats schemes based on these choices, and then we present some examples.

Let us consider the stencil of the general \ats scheme in \req{general_deriv_at2}, which is given by the limits $\smin$ and $\smax$ in space and $\lmin{j}$ and $\lmax{j}$ in time.
If $\smin=\smax$, the number of points are equal on either sides of the point of computation $i$. 
We refer to this as a {\em symmetric} stencil in space.
Else, $\smin \ne \smax$ and the stencil is {\em asymmetric}.
% uniform delay
Regarding the nature of the delays, schemes can potentially have different delay values at different points in a stencil.
However, enforcing a uniform delay across all the buffer points in a scheme, that is $\k{i+j}=\k{}$ for all $i+j\in B$, may lead to linearly dependent conditions and a simpler implementation of schemes.
We can, thus, classify schemes according to the presence or absence of uniform delay in schemes.
% artificial delay
In addition to the uniformity of delays, schemes can also be classified with respect to the time levels chosen at interior points. 
The function at these points can be either at the synchronous time level or at artificially imposed levels which, as we have shown, provide some numerical advantages.

% nature of coefficients
Schemes can also be classified on the basis of the nature of coefficients.
When asynchrony is present, a stencil with symmetric points may not necessarily give rise to symmetry in its coefficients.
This is due to the non-uniform time levels in the stencil at these points.
To obtain symmetry in coefficients, as in standard central difference schemes, we have earlier proposed to use a sum or difference of the function at symmetric grid points.
%Note that coefficients in a scheme can be a function of the random delay $\k{j}$, as seen examples 1 and 3.
In this regard, we classify schemes with symmetric coefficients as the ones which have $|\co{-j}{0}|= |\co{j}{0}|$ (i.e. when $\k{i+j}=0$).
A summary of these classifications is given in \rtab{class}. 

% table
\begin{table}[h]
\begin{center}
\begin{tabular}{l|l l}
\hline
Feature & Classification  \\
\hline
\hline
Layout of & symmetric & asymmetric \\
grid points & $\smin=\smax$ & $\smin \neq \smax$  \\
\hline
Nature of delay &  unconstrained  & uniform delay \\
at buffer points &    & $\k{i+j} = \tilde{K} \ \ \forall \ \ i+j\in B$ \\
\hline
Artificial delay & zero delay & non-zero delay \\
at interior point & $\k{i+j}=0 \ \ \forall \ \ i+j\in I$ & $\k{i+j}\le 0 \ \ \forall \ \ i+j\in I$ \\
\hline
Coefficients & symmetric &  asymmetric  \\
 & $|\co{-j}{0}|= |\co{j}{0}|$ & $|\co{-j}{0}|\neq |\co{j}{0}|$  \\
\hline
\end{tabular}
\caption{Summary of classification of asynchrony-tolerant (AT) schemes.}
\label{tab:class}
\end{center}
\end{table}

From the discussions in previous sections, it is clear that each of these classifications will have consequences in terms of numerical properties and computational performance of schemes.
We will now proceed to derive three \ats schemes and demonstrate how the conditions corresponding to the classification can be implemented in arriving at them. 
%We now proceed to derive four asynchrony-tolerant schemes that are representative of different classes of schemes.

\vspace{0.25cm}
% example 1 and 2
\noindent{\em{Example 1}}: first derivative - second order accurate ($d=1$, $a=2$) \newline
%Let us approximate the first derivative which is second order accurate in space. 
Using $r=2$, 
conditions in \req{cond} are imposed on terms that satisfy the inequality $\pp+2\qq<3$.
This gives rise to a linear system with four equations corresponding to the terms with $(\pp,\qq)=\{(0,0),(1,0),(2,0),(0,1)\}$ in the Taylor series.
The next step is to select a stencil with the function defined at four different combinations of points and time levels. % to obtain the scheme.
%If we assume a non-uniform delay on the right of stencil, that is $i+j\in B$ for $j>0$, then the latest available time level at these points is given by $\k{i+j}$.
Let, as before, $n-\k{i+j}$ be the latest available time level at a point $i+j \in B$ with $j>0$.
Further, let us choose the function set $\{\U{i-1}{n},\U{i}{n},\U{i+1}{n-\k{i+1}}, \U{i+2}{n-\k{i+2}}\}$ to construct the linear system $\bs{A}\bs{\tilde{c}}=\bs{b}$.
This results in 
\be
  \left[
  \begin{array}{cccc}
  1 & 1 & 1 & 1 \\
  -\dx & 0 & \dx & 2\dx \\
  \frac{\dx^2}{2} & 0 & \frac{\dx^2}{2} & 2\dx^2 \\
  0 & 0 & -\k{i+1}\dt & -\k{i+2}\dt 
  \end{array}
  \right] 
\left[
  \begin{array}{c}
  \co{i-1}{0} \\
  \co{i}{0} \\
  \co{i+1}{\k{i+1}} \\
  \co{i+2}{\k{i+2}} \\
  \end{array}
  \right] 
= \left[
  \begin{array}{c}
  0 \\
  1 \\
  0 \\
  0 \\
  \end{array}
  \right]. 
\label{eq:lsx1}
\ee
The rank of the coefficient matrix in the above equation is $4$, which is equal to the size of the system.
Thus, the choice of stencil results in a scheme without any further adjustments.
After solving for the coefficients, we obtain the scheme as
\bea
\left. \frac{\pd u}{\pd x} \right|_i^n &  = & \frac{(-4\k{i+1}+\k{i+2})\U{i-1}{n}+3\k{i+1}\U{i}{n}-\k{i+2}\U{i+1}{n-\k{i+1}}+\k{i+1}\U{i+2}{n-\k{i+2}}}{2(3\k{i+1}-\k{i+2})\dx} \nonumber \\
 & & + \ord\left( \frac{6\k{i+1}-\k{i+2}}{18\k{i+1}-6\k{i+2}}\dx^2,-\frac{\k{i+1}\k{i+2}}{3\k{i+1}-\k{i+2}}\dt\right). 
\label{eq:ex1-0}
\eea
Note that the coefficients are a function of the random delay at buffer points.
%This scheme has limitations under two circumstances. 
It is easy to see by inspection that this scheme has to be complemented in two specific circumstances.
First, when $3\k{i+1}-\k{i+2}=0$ the approximation has an infinite value.
This can be avoided by artificially altering the delays such that $3\k{i+1}-\k{i+2}\ne0$. 
Second, when the function at both buffer points is at a synchronous time level,
i.e., no delay, the above scheme will result in an indeterminate form.
%However, this issue can resolved by approximating the derivative using
In that case one can use
\be
\left. \frac{\pd u}{\pd x} \right|_i^n = \frac{-3\U{i-1}{n}+3\U{i}{n}-\U{i+1}{n-\k{}}+\U{i+2}{n-\k{}}}{4\dx}+ \ord\left( \dx^2,\k{}\dt \right),
\label{eq:ex1-1}
\ee 
which is obtained by substituting $\k{i+1}=\k{i+2}=\k{}$ and simplifying 
\req{ex1-0}.
It is interesting to see that
the coefficients in the above scheme, with a uniform delay across the buffer points, are independent of the delay value, which eliminates the limitations of the scheme in \req{ex1-0}.
Similar schemes can be derived by considering delays on the left of the stencil.

\vspace{0.25cm}
\noindent{\em{Example 2}}: second derivative - fourth order accurate ($d=2$, $a=4$) \newline
%In this example, we derive a second derivative, $d=2$, that is accurate to an order $a=2$.
The relationship between the time step and grid spacing is 
assumed as $\dt\sim\dx$, that is $r=1$.
For this set of parameters, \req{cond} enforces conditions on 21 terms in the Taylor series, which are highlighted in red in \rfig{ts-ex2}. 
\begin{figure}[h]
\begin{center}
{\ss
\begin{TAB}(e,1cm,1cm){:c:c:c:c:c:c:c:}{:c:c:c:c:c:c:c:}
\bmpc${\colr u}$\empc    &     {\colr\bmpc ${u^{(0,1)}}$ ${ \ldt}$\empc}     &     {\colr\bmpc$u^{(0,2)}$ $\ldt^2$\empc}     &     {\colr\bmpc$u^{(0,3)}$ $ \ldt^3$\empc}    &     {\colr\bmpc$u^{(0,4)}$ $ \ldt^4$\empc}&     {\colr\bmpc$u^{(0,5)}$ $ \ldt^5$\empc}&    {$\dots$} \\
{\colr \bmpc${u}^{(1,0)}$ $\ldx$\empc }     &     {\colr \bmpc${u^{(1,1)}}$  ${\ldx}$\\ ${\ldt}$ \empc}      &    {\colr \bmpc$u^{(1,2)}$  $\ldx$ \\ $\ldt^2$\empc}      &     {\colr \bmpc$u^{(1,3)}$  $\ldx$ \\ $\ldt^3$\empc}      &     {\colr \bmpc$u^{(1,4)}$  $\ldx$ \\ $\ldt^4$\empc}&     {\bmpc$u^{(1,5)}$  $\ldx$ \\ $\ldt^5$\empc}      &     {$\dots$} \\
{\colr \bmpc${u}^{(2,0)}$ $\ldx^2$\empc }     &     {\colr \bmpc${u^{(2,1)}}$  ${\ldx^2}$\\ ${\ldt}$ \empc}      &    {\colr \bmpc$u^{(2,2)}$  $\ldx^2$ \\ $\ldt^2$\empc}      &     {\colr \bmpc$u^{(2,3)}$  $\ldx^2$ \\ $\ldt^3$\empc}      &     { \bmpc$u^{(2,4)}$  $\ldx^2$ \\ $\ldt^4$\empc}&     { \bmpc$u^{(2,5)}$  $\ldx^2$ \\ $\ldt^5$\empc}      &     { $\dots$} \\
{\colr \bmpc${u}^{(3,0)}$ $\ldx^3$\empc }     &     {\colr \bmpc${u^{(3,1)}}$  ${\ldx^3}$\\ ${\ldt}$ \empc}      &    {\colr \bmpc$u^{(3,2)}$  $\ldx^3$ \\ $\ldt^2$\empc}      &     {\bmpc$u^{(3,3)}$  $\ldx^3$ \\ $\ldt^3$\empc}      &     { \bmpc$u^{(3,4)}$  $\ldx^3$ \\ $\ldt^4$\empc}&     { \bmpc$u^{(3,5)}$  $\ldx^3$ \\ $\ldt^5$\empc}      &     {$\dots$} \\
{\colr \bmpc${u}^{(4,0)}$ $\ldx^4$\empc }     &     {\colr \bmpc${u^{(4,1)}}$  ${\ldx^4}$\\ ${\ldt}$ \empc}      &    {\bmpc$u^{(4,2)}$  $\ldx^4$ \\ $\ldt^2$\empc}      &     {\bmpc$u^{(4,3)}$  $\ldx^4$ \\ $\ldt^3$\empc}      &     {\bmpc$u^{(4,4)}$  $\ldx^4$ \\ $\ldt^4$\empc}&     {\bmpc$u^{(4,5)}$  $\ldx^4$ \\ $\ldt^5$\empc}      &     {$\dots$} \\
{\colr \bmpc${u}^{(5,0)}$ $\ldx^5$\empc }     &     {\bmpc${u^{(5,1)}}$  ${\ldx^5}$\\ ${\ldt}$ \empc}      &    {\bmpc$u^{(5,2)}$  $\ldx^5$ \\ $\ldt^2$\empc}      &     {\bmpc$u^{(5,3)}$  $\ldx^5$ \\ $\ldt^3$\empc}      &     {\bmpc$u^{(5,4)}$  $\ldx^5$ \\ $\ldt^4$\empc}&     {\bmpc$u^{(5,5)}$  $\ldx^5$ \\ $\ldt^5$\empc}      &     {$\dots$} \\
{ $\vdots$ } & { $\vdots$ } & {$\vdots$ } & {$\vdots$ } & {$\vdots$ } & {$\vdots$ } & {$\ddots$ } \\
\end{TAB}
}
\caption{Terms in the Taylor series of $\U{i+j}{n-l}$ illustrated in a matrix format. Constants in each term are omitted for clarity. To obtain the scheme in Example 2 conditions in \req{cond} are imposed on the red color terms.}
\label{fig:ts-ex2}
\end{center}
\end{figure}
The resulting linear system has 21 equations which, in principle, will need the function at 21 combinations of points and time levels.
However, from \req{cond-async1} we can see that a stencil with only two time levels, $n$ for interior points and $n-\k{}$ for buffer points, will lead to linearly dependent equations in the system.
We use this choice of stencil to reduce the size of the linear system.
Choosing the limits $\{\smin,\smax\}=\{5,6\}$ in space and assuming the buffer points are on the right side of the stencil, leads to a smaller linear system with 11 equations that has a unique solution. 
Upon solving the system, the resulting \ats scheme is 
\bea
\left. \frac{\pd^2 u}{\pd x^2} \right|_i^n & = & \frac{1}{12\dx^2}\left[ 35\U{i-5}{n}-164\U{i-4}{n}+294\U{i-3}{n}-236\U{i-2}{n}+71\U{i-1}{n}  -45\U{i+1}{n-\k{}} \right. \nonumber\\
 &  & \left. +225\U{i+2}{n-\k{}}-450\U{i+3}{n-\k{}}+450\U{i+4}{n-\k{}}-225\U{i+5}{n-\k{}}+45\U{i+6}{n-\k{}} \right] \nonumber \\
& & +\ord\left( \dx^4 , \k{} \dx^3 \dt \right).
\label{eq:ex2-1}
\eea 
Note that a single linear system has been used to obtain the above scheme.
% In the next example, we derive a scheme that resembles central difference under the absence of asynchrony, using the alternative approach described in \rsec{alt-app}.

\vspace{0.25cm}
\noindent{\em{Example 3}}: second derivative - fourth order accurate ($d=2$, $a=4$) \newline
%Let us derive a scheme to compute a second derivative that is fourth order accurate in a step by step process.  
In this example, we will use the alternative step-by-step approach described in \rsec{alt-app} to derive an \ats scheme that reduces to a standard central difference scheme in the absence of delays.
If we consider the Taylor series of $u$ at a generic point and time level in a stencil, and assume $\dt\sim\dx^2$, conditions have to be imposed on terms with
\bea
(\pp,\qq) & = & \{(0,0), (1,0), (2,0), (3,0), (4,0), (5,0) \nonumber \\
 & & (0,1), (1,1), (2,1), (3,1), (0,2), (1,2)\} .
\label{eq:set}
\eea
In order to maintain a symmetry in the stencil points and coefficients, we use the sum $\left( \U{i+j}{n} + \U{i-j}{n-l} \right)$ in the first step, which eliminates the terms with $(\pp,\qq)=\{(1,0),(3,0)\}$ upon Taylor series expansion. In the second step, conditions are enforced on the terms that are only a function of $\dx$ or the terms with $\qq=0$ by expanding the stencil in space.
These are the three terms corresponding to $(\pp,\qq)=\{(0,0),(2,0),(4,0)\}$, which result in three equations using the function $\left( \U{i+j}{n} + \U{i-j}{n-l} \right)$ for $j=0,1,2$:
\be
  \left[
  \begin{array}{ccc}
  2 & 2 & 2 \\
  0 & \dx^2 & 4\dx^2 \\
  0 &\frac{\dx^4}{12} & \frac{4\dx^4}{3}  \\
  \end{array}
  \right]
\left[
  \begin{array}{c}
  \cats{0}{l} \\
  \cats{1}{l} \\
  \cats{2}{l} \\
  \end{array}
  \right]
= \left[
  \begin{array}{c}
  0 \\
  1 \\
  0 \\
  \end{array}
  \right]
\label{eq:lsx1}
\ee
After solving the equations we obtain a linear combination of the function that is free from lower order synchronous terms.
We can express the linear combination as 
\be
\left. \frac{\pd^2 u}{\pd x^2} \right|_i^n  =  \frac{-\U{i+2}{n} + 16 \U{i+1}{n} - 30\U{i}{n} + 16\U{i-1}{n-l} - \U{i-2}{n-l}}{12\dx^2} + \ord(\dx^4,l\dt,l\dt/\dx^2).
\label{eq:cd4-async}
\ee
When $l=0$, the terms in the truncation error due to asynchrony disappear from the above expression, and we recover the standard fourth order central difference scheme. 
In the next step, we eliminate the remaining lower order terms which appear due to asynchrony.
% in the time level of $u$, we will arrive at a fourth order asynchrony-tolerant scheme.
If we expand the stencil further in space, which corresponds to $j>2$, then the scheme would possess the required symmetries, but will not reduce to the fourth order central difference in the absence of delay.
On the other hand, when the stencil size is increased in time, i.e., $l\in\{\k{},\k{}+1,\k{}+2,\dots\}$, we get a scheme that does resemble a standard central difference. 
The conditions on the six asynchrony terms from the set in \req{set}, need \req{cd4-async} at six time levels. 
However, with the use of multiple time levels in the stencil, 
the resulting linear system has three linearly dependent conditions.
We find that conditions on terms with the same $\qq$ are all mathematically equivalent.
This reduces the size of the linear system that uses the linear combination in \req{cd4-async} at $l\in\{\k{},\k{}+1,\k{}+2\}$ to three equations:
\be
  \left[
  \begin{array}{ccc}
  -\k{}\frac{5\dt}{4\dx^2} & -(\k{}+1)\frac{5\dt}{4\dx^2} & -(\k{}+2)\frac{5\dt}{4\dx^2} \\
 \k{}^2 \frac{5\dt^2}{8\dx^2} & (\k{}+1)^2\frac{5\dt^2}{8\dx^2} & (\k{}+2)^2\frac{5\dt^2}{8\dx^2} \\
  1 & 1 & 1  \\
  \end{array}
  \right]
\left[
  \begin{array}{c}
  \cats{0}{\k{}} \\
  \cats{1}{\k{}+1} \\
  \cats{2}{\k{}+2} \\
  \end{array}
  \right]
= \left[
  \begin{array}{c}
  0 \\
  0 \\
  1 \\
  \end{array}
  \right]
\label{eq:lsx1}
\ee
The solution to this linear system results in the scheme

\bea
\left. \frac{\pd^2 u}{\pd x^2} \right|_i^n &  = & \frac{1}{2}(\k{}^2+3\k{}+2)\frac{-\U{i+2}{n} + 16 \U{i+1}{n} - 30\U{i}{n} + 16\U{i-1}{n-\k{}} - \U{i-2}{n-\k{}}}{12\dx^2}  \nonumber \\
 & & -(\k{}^2+2\k{})\frac{-\U{i+2}{n} + 16 \U{i+1}{n}- 30\U{i}{n} + 16 \U{i-1}{n-\k{}-1} - \U{i-2}{n-\k{}-1}}{12\dx^2} \nonumber \\
 & & +\frac{1}{2}(\k{}^2+\k{})\frac{-\U{i+2}{n} + 16 \U{i+1}{n}- 30\U{i}{n} + 16 \U{i-1}{n-\k{}-2} - \U{i-2}{n-\k{}-2}}{12\dx^2} \nonumber \\
 & & + \ord\left(\dx^4,\k{}(\k{}+1)(\k{}+2)\dt^3,\k{}(\k{}+1)(\k{}+2) \dt^3/\dx^2 \right).
\label{eq:cd4-at}
\eea
Clearly, in the absence of delay, $\k{}=0$, the scheme reduces to a standard fourth order central difference scheme.
Note that, like the scheme in Example 1, the coefficients in the above scheme are a function of the random delay $\k{}$.
However, unlike \req{ex1-0} in Example 1, this scheme can take any delay value in the range $[0,L-1]$. 
%in this case there are no limitations (like resulting in an indeterminate form) in the implementation of the scheme. 

\ \\

Some other useful examples with their leading 
order term in the truncation error are collected in 
\rtabs{atschemes-left}{atschemes-right}.
These will be used later on 
when we assess the numerical performance of 
different \ats schemes.

\section{Error analysis}
\label{sec:error}
In previous sections, we presented a method to derive \ats schemes of arbitrary accuracy.
As explained earlier, these schemes are, typically, used at PE boundaries ($i\in\Ib$) where asynchrony is experienced.
The number of computations that are carried out asynchronously in a domain depend on the number of PEs used to solve the problem, the stencil size of schemes used at interior points and statistics of the random delays, which in turn depend on the characteristics of communications in a computing system.
%The asynchronous computing method poses two issues in understanding the overall accuracy of a numerical solution.
These dependencies bring new challenges when trying to understand the overall accuracy of these \ats schemes.
First, due to the random nature of the delay, the associated truncation error is also random in nature.
Second, schemes to compute spatial derivatives at interior points are not the same as \ats schemes at PE boundary points and have, thus, different truncation errors. These issues result in a non-homogeneity of error in the domain, both in space as well as time.

In our previous work \cite{DA2014}, we have proposed a statistical description to analyze the overall error and determine the accuracy of the numerical solution.
We follow a similar procedure in this work. 
Before we develop the error analysis, we present some necessary definitions that will be used. 
First, let us define the probability of having a time level $\n=n-\k{i}$ at a grid point $i$ as $\prob{k}{i}$.
The sum of probabilities of all levels at point $i$ is obviously
\begin{equation}
\sum_{k[i]=0}^{L-1} \prob{k}{i} = 1.
\label{eq:c6}
\end{equation}
To obtain the statistics of the error, we define two types of averages for a variable $f$: a space average and an ensemble average.
The space average can be performed over all points in the set $I$ or the subsets $\Ii$ and $\Ib$.
If the average is over the entire domain, that is $i\in I$, it is denoted by angular brackets and given by $\la f\ra = \sum_{i=1,N} f_i / N$.  
On the other hand, the average over the points in the subsets $\Ii$ and $\Ib$ are given by $\xaveb{f}= \sum_{i\in \Ib} f_i / \Nb$ and
$\xavei{f} = \sum_{i\in \Ii} f_i / \Ni$,
respectively.
The random nature of delays is taken into account by ensemble averages, which is denoted by an overline $\eave{f}$.

%fde and error
A common measure of the error incurred by using a finite difference representation of the original PDE is given by the so called truncation error. Formally, it is given by the difference between the PDE and the approximate finite difference equation (FDE), that is $E=PDE-FDE$.
As introduced in \cite{DA2014}, the assessment of the error of asynchrony schemes which are random in nature and heterogeneous in space, can be done by applying the two averages described above. That is,
%The overall error in the domain can be expressed as the average of error across all the points.
%Using the space and ensemble averages, the overall error in the domain can be computed from
\be
\ave{E} = {1\over N}\sum_{i=1,N} \eave{\E{i}{n}},
\ee
where $\E{i}{n}$ is the truncation error at the point $i$ and time level $n$.
Due to the non-uniform expression for the truncation error at interior and PE boundary points, it is convenient to split the error according to the two sets of points:
\be
\ave{E} = {1\over N}\left[ \sum_{i\in\Ii} \E{i}{n} + \sum_{i\in\Ib} \eave{\ranE{i}{n}} \right]
\label{eq:aveE_split}
\ee
Note that the error due to interior points does not possess randomness due to delays and are, hence, unaffected by the ensemble average.
On the other hand, errors at PE boundary points have both random asynchronous and deterministic synchronous components.
%If we consider the error $\E{i}{n}$, it is the sum of truncation error due the approximation of each term in PDE.
This allows us to further split the error in the set $\Ib$ as 
\be
\ave{E} = {1\over N}\left[ \sum_{i\in\Ii} \E{i}{n} + \sum_{i\in\Ib} {\E{i}{n}}|_s + \sum_{i\in\Ib} \eave{\ranE{i}{n}}|_a \right],
\label{eq:aveE_split2}
\ee
where the subscripts $s$ and $a$ denote the synchronous and asynchronous components, respectively.
It is clear that in the absence of delays $\eave{\ranE{i}{n}}|_a = 0$.

The order of accuracy of a scheme will depend on the leading order term in each of the error terms in the above equation.
%These terms are sum of truncation error due approximation of each derivative in PDE, which include time derivatives.
These terms comprise the sum of the truncation error due to all the
derivatives in the original PDE, including the time derivative.
%The leading order term can be from any of the derivatives. 
Thus, it is important to choose the accuracy of time integration to match the order of accuracy of space derivatives. 
We will discuss this topic next and then present an example to illustrate the effect of asynchrony on the error.
We will end this section with a generalization of the results on accuracy of asynchronous schemes. 

\subsection{Time integration}
\label{sec:time-disc}
To understand the effect of time discretization on the overall order of accuracy, let us consider the equation $\pd u/\pd t = f$, where $f$ depends on spatial derivatives of $u$, integrated using Euler scheme. 
%start with an example of Euler scheme to solve $\pd u/\pd t = f$, where $f$ is a function of spatial derivatives.
The scheme is first order in time with the leading order term being $-u^{(0,2)}\dt / 2$.
As mentioned in \rsec{at-gm}, if we assume a relation of the form $\dt\sim \dx^r$, then the leading order term is equivalent to $\ord(\dx^r)$ in space.
When the accuracy of the space derivatives is greater than $r$, the total error will, very likely, be dominated by the temporal term and will dictate the order of accuracy of the solution. 

Thus, if a certain order is desired for space derivatives, it is important to select a time discretization with the same (or greater) order to keep the overall order unchanged.
We will follow this practice as we demonstrate the accuracy of the proposed \ats schemes next.
%To demonstrate the accuracy of asynchrony-tolerant schemes using numerical experiments, it is important to choose a time discretization scheme such that its error does not dominate total error. 
%In other words, the equivalent order of accuracy of the time scheme should equal or more than the asynchrony-tolerant schemes.
For this, we choose a linear multi-step method to compute the time derivative. A general expression with $T$ time steps is given by 
\be
u^{n+1}_i=u^{n}_i+ \dt \sum_{m=0}^{T-1} \beta_m f_i^{n-m},
\label{eq:time_disc}
\ee
where the coefficients $\beta_m$ determine the particular temporal scheme \cite{Stoer2013}.

The advantage of using a temporal scheme of the form \req{time_disc} is that the terms $f_i^{n-m}$ can be computed using \ats schemes and are thus, free of asynchrony errors to the desired order of accuracy. Thus, so will be the linear combination of $f_i$ at different time steps.
For example, if one uses an \ats scheme that is fourth order accurate, with $r=2$ (i.e. $\dt\sim \dx^2$), then one needs a temporal scheme with second order accuracy to maintain fourth order accuracy globally.
This can be accomplished by a two-step Adams-Bashforth method 
\be
u^{n+1}_i=u^{n}_i+ \dt \left(\frac{3}{2} f_i^n - \frac{1}{2} f_i^{n-1} \right),
\label{eq:time_ab2}
\ee
which is readily shown to be second order in time \cite{Stoer2013}. The generalization to higher orders is straightforward.

\subsection{Example: heat equation with fourth-order accurate \ats schemes}
Let us consider the 1D heat equation, 
\begin{equation}
\frac{\pd u}{\pd t}=\alpha\frac{\pdd u}{\pd x^2},
\label{eq:heat}
\end{equation}
where $u(x,t)$ is the temperature and $\alpha$ is the thermal diffusivity of the medium.
The above equation is solved on a uniform grid shown in \rfig{grid} with periodic boundary conditions.
The equation is approximated with the second order Adams-Bashforth scheme shown in \req{time_ab2} and standard fourth order central difference for the space derivative at interior points.
At the PE boundary points, the space derivative is computed with a fourth
order \ats scheme which, with delay in the left boundary, 
is given by \req{cd4-at} derived in Example 3.

Using Taylor series, the truncation error at interior points is
\be
\E{i}{n} = \left( -\frac{1}{6} \U{}{(0,3)} -  \frac{1}{4} \alpha \U{}{(2,2)}  \right) \dt^2 - \frac{1}{90} \alpha \U{}{(6,0)} \dx^4 + \ord \left( \dx^6,\dt^3,\dx^4\dt \right) .
\label{eq:te-sync}
\ee
As mentioned above, at PE boundary points, the truncation error can be split into the synchronous and asynchronous components,
\be
\ranE{i}{n}|_{\k{}=k} = {\E{i}{n}}|_s + \ranE{i}{n}|_{a,\k{}=k}.
\label{eq:aveE_split3}
\ee
Because by construction, the \ats scheme in \req{cd4-at} reduces to the standard central difference in the absence of delay, the synchronous component of the error, ${\E{i}{n}}|_s$,
 is the same as \req{te-sync}. Note that since this scheme has 
 a uniform delay
 at buffer points, we drop the subscript for $\k{}$ in the above
 expression for simplicity.
The asynchronous component of the error, considering delays only on the left side of the stencil, can be readily shown to be  
%\bea
%\ranE{i}{n}|_{a,\k{}=k} & = & \left( -\frac{1}{6} \U{}{(0,3)} -  \frac{1}{4} \alpha \U{}{(2,2)}  \right) \dt^2 - \frac{1}{90} \alpha \U{}{(6,0)} \dx^4 \nonumber \\
%& & - \frac{5}{24}\left( k^3 + 3 k^2 + 2 k \right) \alpha \U{}{(0,3)} \frac{\dt^3}{\dx^2}  + \ord \left( \dx^6,\dt^3,\dt^3/\dx \right), 
%\label{fig:te-at}
%\eea
\be
\ranE{i}{n}|_{a,\k{}=k} =  - \frac{5}{24}\left( k^3 + 3 k^2 + 2 k \right) \alpha \U{}{(0,3)} \frac{\dt^3}{\dx^2} + \ord\left( k^3\dt^3/\dx \right).
\label{fig:te-async}
\ee
The leading order term in the error remains the same when the delays are experienced on the right of the stencil.
Clearly, when $\k{}=0$, we have $\ranE{i}{n}|_{a,\k{}=k}=0$ and thus also its ensemble average.
On the other hand, if $\k{}\ge0$, then the ensemble average is
\bea
\eave{\ranE{i}{n}|_{a}}
& \ap&
      \sum_{k=0}^{L-1} p_k{\ranE{i}{n}|_{a,\k{}=k}} \nonumber \\
       %\nonumber \\
& \ap&
      \sum_{k=0}^{L-1} p_{k}\left( - \frac{5}{24}\left( k^3 + 3 k^2 + 2 k \right) \alpha \U{}{(0,3)} \frac{\dt^3}{\dx^2} \right) \nonumber \\
& \ap&
      \left( - \frac{5}{24} \alpha \U{}{(0,3)} \frac{\dt^3}{\dx^2} \right)
       \sum_{k=0}^{L-1} p_{k}\left( k^3 + 3 k^2 + 2 k \right)  \nonumber \\
& \ap&
      \left( - \frac{5}{24} \alpha \U{}{(0,3)} \frac{\dt^3}{\dx^2} \right)
       \left( \kmom{3} + 3 \kmom{2} + 2 \kmom{} \right), 
\label{eq:te-async1}
\eea
where moments are given by $\eave{\k{}^n}=\sum_{k=0,L-1} p_{k}k^n$.
It is interesting that the average error under the presence of asynchrony, depends not just on the mean of the delay as in \cite{DA2014}, but also on its higher order moments.
The implication of this result is that in assessing the performance of asynchronous numerical schemes a certain degree of details about the architecture of the computing system would be needed, such as the probability density function of the delays $\k{}$. Conversely, one can quantitatively compare the performance of different computing systems by comparing moments of $\k{}$.

%In the next step, the space average of the error in the domain can be  computed by substituting 
We now substitute the leading order terms in \reqs{te-sync}{te-async1} into \req{aveE_split3}. Assuming the statistics of the delays are homogeneous in space, the average error is
\bea
\ave{E} & \ap & {1\over N}\left[ \sum_{i\in\Ii} \left( \left( -\frac{1}{6} \U{}{(0,3)} -  \frac{1}{4} \alpha \U{}{(2,2)}  \right) \dt^2 - \frac{1}{90} \alpha \U{}{(6,0)} \dx^4 \right) \right. \nonumber \\
 & & + \sum_{i\in\Ib} \left( \left( -\frac{1}{6} \U{}{(0,3)} -  \frac{1}{4} \alpha \U{}{(2,2)}  \right) \dt^2 - \frac{1}{90} \alpha \U{}{(6,0)} \dx^4 \right) \nonumber \\
 & & \left. + \sum_{i\in\Ib} \left( \left( - \frac{5}{24} \alpha \U{}{(0,3)} \frac{\dt^3}{\dx^2} \right)  \left( \kmom{3} + 3 \kmom{2} + 2 \kmom{} \right) \right) \right].
\label{eq:aveE_split4}
\eea
The first two sums on the right hand side are due to synchronous
computations and can be conveniently combined by noting that $I=I_I\cup\Ib$. 
%As expressions of these two components are the same, we can use the summation over all the point in $I$, instead of the subsets $I_I$ and $I_B$.
To determine the spatial accuracy of the solution, we use the stability parameter $r_\alpha = \alpha \dt / \dx^2$ to substitute the time step $\dt$ in terms of $\dx$.
This corresponds to $r=2$ in the formulation presented in \rsec{at-gm}. The above equation then reduces to 
\bea
\ave{E} & \ap & {1\over N}\left[ \sum_{i\in I} \left( -\frac{1}{6} \frac{r_\alpha^2}{\alpha^2} \U{}{(0,3)} -  \frac{1}{4} \frac{ r_\alpha^2}{\alpha} \U{}{(2,2)} - \frac{1}{90} \alpha \U{}{(6,0)}\right) \dx^4 \right. \nonumber \\
 & & \left. + \sum_{i\in\Ib} \left( - \frac{5}{24} \frac{ r_\alpha^3}{\alpha^2} \U{}{(0,3)} \right)  \left( \kmom{3} + 3 \kmom{2} + 2 \kmom{} \right) \dx^4 \right],
\label{eq:aveE_split5}
\eea
%Using the space averages defined earlier in this section, we can now write 
which can be rewritten as
\bea
\ave{E} & \ap &  \left[ -\frac{1}{6} \frac{ r_\alpha^2}{\alpha^2} \la\U{}{(0,3)}\ra -  \frac{1}{4} \frac{ r_\alpha^2}{\alpha} \la \U{}{(2,2)} \ra - \frac{1}{90} \alpha \la\U{}{(6,0)}\ra \right] \dx^4 \nonumber \\
 & & + \left[\frac{N_B}{N} \left( \kmom{3} + 3 \kmom{2} + 2 \kmom{} \right) \left( - \frac{5}{24} \frac{ r_\alpha^3}{\alpha^2} \la \U{}{(0,3)} \ra_B \right) \right] \dx^4. 
\label{eq:aveE_split6}
\eea
The average error is clearly seen to possess components due to synchronous and asynchronous computations.
Either of the terms can dominate the average error depending on physical
parameters ($\alpha$, initial conditions, etc.), numerical parameters 
($\dx$, $r_\alpha$, etc.), and simulation parameters ($P$, network
performance, etc.).
If the synchronous part dominates the overall error, then the resulting
scheme is fourth order accurate, that is $\ave{E}\sim \ord(\dx^4)$.

If, on the other hand, the asynchronous component dominates, the error is given by
\be
\ave{E} \ap  \frac{P(\smin+\smax)}{N} \left( \kmom{3} + 3 \kmom{2} + 2 \kmom{} \right) \left( - \frac{5}{24} \frac{r_\alpha^3}{\alpha^2} \la \U{}{(0,3)} \ra_B \right) \dx^4, 
\label{eq:aveE_2}
\ee 
where we have used $N_B=(\smin+\smax)P$, with $\smin$ and $\smax$ being the stencil size in space at interior points.
Using $N=\mathcal{L}/\dx$, where $\mathcal{L}$ is the length of the domain, and for all other parameters kept constant, the average error is found to scale as
\bea
\ave{E} & \sim & \frac{P}{N} \left( \kmom{3} + 3 \kmom{2} + 2 \kmom{} \right) \dx^4 \nonumber \\
& \sim & {P}\left( \kmom{3} + 3 \kmom{2} + 2 \kmom{} \right) \dx^5
\label{eq:aveE_scaling}
\eea
Interestingly, the order of accuracy of the numerical method now depends on how the problem is scaled on a parallel machine.
In the case of weak scaling, where the computational effort per PE is kept constant, that is $P/N=constant$, the error varies as $\dx^4$ and the method is fourth order accurate in space.
On the other hand, when the total computational effort is kept constant ($N=constant$) and the simulations are carried out on increasingly large number of PEs, the average error is $\ave{E}\sim \ord(\dx^5)$ and the method is fifth order accurate.
We also observe that the error scales linearly with $P$.

In some situations the error due the synchronous and 
asynchronous components may be
comparable. In such cases, the overall error will 
%have a behavior that is
%dictated by the sign (positive or negative) of the leading order terms in
%the error of each component. 
depend on the sign of each contribution. 
If the synchronous and asynchronous components
have opposite signs, then it is possible to expect some error 
cancellation.
%the error in asynchronous computing will be lower
%that the corresponding synchronous computing method, although, 
The order of accuracy though would remain unaltered. 

\subsection{Generalization}
We now proceed to generalize the expressions for average error ($\ave{E}$) presented in \req{aveE_scaling}.
For this, we restate the conditions and assumptions that lead to \req{aveE_scaling}.
First, we assumed that the asynchronous component dominates the overall error.
Second, the \ats scheme used in the analysis is fourth order accurate, which lead to an $\ord(\dx^4)$ leading term due to asynchrony.
We have also assumed a uniform random delay in the stencil, that is $\k{i+j}=\k{}$ for all $i+j \in B$.
Also, the scheme uses three successive asynchronous time levels (with delays $\k{}$, $\k{}+1$, $\k{}+2$), which results in a cubic polynomial in $k$ in the leading order error term.

With the above observations, we can arrive at a general case which uses \ats schemes with $\mathcal{T}$ number of successive asynchrony time levels and is accurate to an order $a$. 
If the asynchronous component of the error 
dominates the average error, then it is easy to
generalize \req{aveE_scaling} as:
\bea
\ave{E} & \sim & \frac{P}{N} \dx^a \sum_{m=1}^{\mathcal{T}}\gamma_m\kmom{m} \nonumber \\
& \sim & {P}\dx^{a+1} \sum_{m=1}^{\mathcal{T}}\gamma_m\kmom{m}
\label{eq:aveE_scaling_gen}
\eea
Note that the average error still scales linearly with the number of PEs.
However, higher order moments of the delay are necessary to characterize the error when the stencil size of \ats schemes is expanded in time.
A minimum accuracy of order $a$ is then assured, regardless of how simulations are scaled up. 
 
\section{Numerical Simulations}
\label{sec:simulations}
In this section, we verify the numerical performance of AT schemes.
Let us consider the general PDE:
\be
{\partial u\over \partial t} = \sum_{d=1,\cal{D}} \beta_d {\partial^d u \over
\partial x^d}
\label{eq:general}
\ee 
where $\cal{D}$ is the highest derivative and the coefficient $\beta_d$
determines the characteristics of the physical process associated with the
$d$-th  derivative. 
Of particular interest are the heat equation (${\cal D}=2$ with $\beta_1=0$ and
$\beta_2=\alpha$), and the advection-diffusion equation (${\cal D}=2$ with
$\beta_1=c$ and $\beta_2=\alpha$), where $\alpha$ is the thermal or viscous
diffusivity and $c$ is the advection speed.
When the coefficients $\beta_d$ are constant, \req{general} 
is linear and usually
possesses an analytical solution, which will be used here to evaluate the error in
numerically computed solutions. 
The so-called nonlinear viscous Burgers' equation, which is widely used in
understanding physical properties of fluid flows, is obtained with ${\cal D}=2$,
$\beta_1=u(x,t)$ and $\beta_2=\alpha$. 
We also perform simulations of this equation to demonstrate the feasibility
of AT schemes in solving multi-scale phenomena with non-linear couplings.

\subsection{Simulation details}
The equations described in the above section are solved in a periodic domain of length $2\pi$. For initial conditions, we use a multi-scale spectrum given by superimposed sinusoidal waves:
\be
u(x,0) = \sum_\kappa A(\kappa)\sin(\kappa x + \phik),
\label{eq:IC}
\ee
where $\kappa$ denotes the wavenumber. $A(\kappa)$ and $\phik$ are the
amplitude and phase angle corresponding to each wavenumber $\kappa$.
The phase $\phik$ are included in order to avoid circumstances like
the coincidence of PE boundaries with zero-gradients in the function, which may
result in very special cancellations 
of some of the error terms due to asynchrony. 
The results presented below are in fact 
ensemble averages of multiple simulations with different phases.
In addition to avoiding special cases in terms of accuracy as mentioned
above, this procedure 
provides a probability space over which ensemble averages can be obtained.

% discretization
Simulations are carried out using several configuration cases with different
governing equations to study the behavior of AT schemes in different regimes. 
The synchronous computations of spatial derivatives are carried out using
standard central difference schemes.
Close to PE boundary points, the AT schemes summarized in 
\rtabs{atschemes-left}{atschemes-right} are used.
The time derivatives are discretized according to the procedure described in
\rsec{time-disc}. 
%For second, fourth and sixth order accurate spatial derivatives, we use Euler, second order Adams-Bashforth and third order Adams-Bashforth, respectively.
The details of each numerical experiment 
are tabulated in \rtab{cases}.

% tablr
\begin{table}[h]
\begin{center}
\begin{tabular}{c|c c c c}
\hline
Case & Equation & Time derivative & \multicolumn{2}{c}{Space derivatives}  \\
 & & &  Synchronous & Asynchrony-tolerant \\ 
\hline
\hline
1 & AD & Eul & CD2 & $(1,2,2)b$, $(2,2,2)b$ \\
2 & D & Eul & CD2 & $(2,1,2)$ \\
3 & AD & Eul & CD2 & $(1,2,2)a$, $(2,2,2)a$ \\
4 & AD & AB2 & CD4 & $(1,4,2)$, $(2,4,2)$ \\
5 & D & AB3 & CD6 & $(2,6,2)$ \\
6 & VB & AB2 & CD4 & $(1,4,2)$, $(2,4,2)$ \\
\end{tabular}
\caption{Parameters of numerical experiments. In the table: AD - linear advection-diffusion equation, D - diffusion equation, VB - non-linear viscous Burgers' equation; Eul - first order Euler scheme, AB2 and AB3 - second and third order Adams-Bashforth schemes; CD2, CD4 and CD6 - second, fourth and sixth order central difference schemes; AT schemes referred according to $(d,a,r)$ notation in \rtabs{atschemes-left}{atschemes-right}.}
\label{tab:cases}
\end{center}
\end{table}

% table
\afterpage{%
    \clearpage% Flush earlier floats (otherwise order might not be correct)
    \thispagestyle{empty}% empty page style (?)
\begin{landscape}
\begin{table}[h]
\begin{center}
{\tabulinesep=0.5mm
\begin{tabu}{|c|c|c|}
\hline
Scheme & {Scheme at} & Leading order terms \\
$(d,a,r)$ & left boundary & \\
\hline
\hline
%%%%%%%%%%%%%%%%
$(2,1,2)$
%,  uniform delay 
 & 
{ $
\mysize 
\begin{array} {c}  
%{\tiny $\begin{array} {c}  
%\frac{\U{i+1}{n}-\U{i}{n}-\U{i-1}{n-\k{}}+\U{i-2}{n-\k{}}}{2\dx^2}
\left({\U{i+1}{n}-\U{i}{n}-\U{i-1}{n-\k{}}+\U{i-2}{n-\k{}}}\right)/{2\dx^2}
\end{array}$}
&
{ $
\mysize
\begin{array} {c}  
%{\tiny $\begin{array} {c}  
\frac{1}{2} \k{} u^{(1,1)}\frac{\dt}{\dx},
\frac{1}{2} u^{(3,0)}{\dx}
\end{array}$}
  \\
\hline
%%%%%%%%%%%%%%%%
$(1,2,2)a$ & %, uniform delay & 
{ $\mysize \begin{array} {c}  
%{\tiny $\begin{array} {c}  
%\frac{3\U{i+1}{n}-3\U{i}{n}-\U{i-1}{n-\k{}}-\U{i-2}{n-\k{}}}{4\dx}
\left({3\U{i+1}{n}-3\U{i}{n}-\U{i-1}{n-\k{}}-\U{i-2}{n-\k{}}}\right)/{4\dx}
\end{array}$}
&
{ $\mysize \begin{array} {c}  
%{\tiny $\begin{array} {c}  
\frac{1}{4} {\k{}} u^{(1,1)}\dt,
\frac{5}{12} u^{(3,0)}{\dx^2}
\end{array}$}
  \\
\hline
%%%%%%%%%%%%%%%%
$(2,2,2)a$ & %, uniform delay & 
{ $\mysize \begin{array} {c}  
%{\tiny $\begin{array} {c}  
\left({2\U{i+2}{n}-4\U{i+1}{n}+2\U{i}{n}+\U{i-1}{n-\k{}}-2\U{i-2}{n-\k{}}+\U{i-3}{n-\k{}}}\right)/{3\dx^2}
\end{array}$}
&
{ $\mysize \begin{array} {c}  
%{\tiny $\begin{array} {c}  
\frac{1}{3} {\k{}} u^{(2,1)}\dt,
\frac{13}{12} u^{(4,0)}{\dx^2}
\end{array}$}
  \\
\hline
%%%%%%%%%%%%%%%%%
$(1,2,2)b$ & %, uniform delay &
{ $\mysize \begin{array} {c}  
%{\tiny $\begin{array} {c}  
\left({\U{i+1}{n}-(\k{}+1)\U{i-1}{n-\k{}}+\k{}\U{i-1}{n-\k{}-1}}\right)/{2\dx}
\end{array}$}
&
{ $\mysize \begin{array} {c}  
%{\tiny $\begin{array} {c}  
\frac{1}{6} u^{(3,0)}{\dx^2}
\end{array}$}
  \\
\hline
%%%%%%%%%%%%%%%%
$(2,2,2)b$ & %, uniform delay &
{ $\mysize \begin{array} {c}  
%{\tiny $\begin{array} {c}  
\left({\U{i+1}{n}-2\U{i}{n}+(\k{}+1)\U{i-1}{n-\k{}}-\k{}\U{i-1}{n-\k{}-1}}\right)/{\dx^2}
\end{array}$}
&
{ $\mysize \begin{array} {c}  
%{\tiny $\begin{array} {c}  
\frac{1}{2} {\k{}(\k{}+1)} u^{(0,2)}\frac{\dt^2}{\dx^2},
\frac{1}{12} u^{(4,0)}{\dx^2}
\end{array}$}
  \\
\hline
%%%%%%%%%%%%%%%%
$(1,4,2)$ & %, uniform delay &
{ $\mysize \begin{array} {c}  
%{\tiny $\begin{array} {c}  
\frac{1}{2}(\k{}^2+3\k{}+2)\left({-\U{i+2}{n} + 8 \U{i+1}{n} - 8\U{i-1}{n-\k{}} + \U{i-2}{n-\k{}}}\right)/{12\dx}  \\
 -(\k{}^2+2\k{})\left({-\U{i+2}{n} + 8 \U{i+1}{n} - 8 \U{i-1}{n-\k{}-1} + \U{i-2}{n-\k{}-1}}\right)/{12\dx} \\
 +\frac{1}{2}(\k{}^2+\k{})\left({-\U{i+2}{n} + 8 \U{i+1}{n} - 8 \U{i-1}{n-\k{}-2} + \U{i-2}{n-\k{}-2}}\right)/{12\dx}
\end{array}$}
&
{ $\mysize \begin{array} {c}  
%{\tiny $\begin{array} {c}  
\frac{1}{30} u^{(5,0)}{\dx^4}
\end{array}$}
  \\
\hline
%%%%%%%%%%%%%%%%
$(2,4,2)$ & %, uniform delay &
{ $\mysize \begin{array} {c}  
%{\tiny $\begin{array} {c}  
\frac{1}{2}(\k{}^2+3\k{}+2)\left({-\U{i+2}{n} + 16 \U{i+1}{n} - 30 \U{i}{n} + 16\U{i-1}{n-\k{}} - \U{i-2}{n-\k{}}}\right)/{12\dx^2}  \\
 -(\k{}^2+2\k{}) \left({-\U{i+2}{n} + 16 \U{i+1}{n} - 30 \U{i}{n} + 16\U{i-1}{n-\k{}-1} - \U{i-2}{n-\k{}-1}}\right)/{12\dx^2} \\
 +\frac{1}{2}(\k{}^2+\k{}) \left({-\U{i+2}{n} + 16 \U{i+1}{n} - 30 \U{i}{n} + 16\U{i-1}{n-\k{}-2} - \U{i-2}{n-\k{}-2}}\right)/{12\dx^2} 
\end{array}$}
&
{ $\mysize \begin{array} {c}  
%{\tiny $\begin{array} {c}  
\frac{5}{24} {\k{}(\k{}+1)(\k{}+2)} u^{(0,3)}\frac{\dt^3}{\dx^2}, \\
\frac{1}{90} u^{(6,0)}{\dx^4}
\end{array}$}
  \\
\hline
%%%%%%%%%%%%%%%%%
$(2,6,2)$ & %, uniform delay &
{ $\mysize \begin{array} {c}  
%{\tiny $\begin{array} {c}  
\frac{1}{6}(\k{}^3+6\k{}^2+11\k{}+6) \\
\left({2\U{i+3}{n}-27\U{i+2}{n} + 270 \U{i+1}{n} - 490 \U{i}{n} + 270 \U{i-1}{n-\k{}} - 27\U{i-2}{n-\k{}} + 2\U{i-3}{n-\k{}}}\right)/{180\dx^2}  \\
- \frac{1}{2}(\k{}^3+5\k{}^2+6\k{}) \\
\left({2\U{i+3}{n}-27\U{i+2}{n} + 270 \U{i+1}{n} - 490 \U{i}{n} + 270 \U{i-1}{n-\k{}-1} - 27\U{i-2}{n-\k{}-1} + 2\U{i+3}{n-\k{}-1}}\right)/{180\dx^2}  \\
+ \frac{1}{2}(\k{}^3+4\k{}^2+3\k{}) \\
\left({2\U{i+3}{n}-27\U{i+2}{n} + 270 \U{i+1}{n} - 490 \U{i}{n} + 270 \U{i-1}{n-\k{}-2} - 27\U{i-2}{n-\k{}-2} + 2\U{i+3}{n-\k{}-2}}\right)/{180\dx^2}  \\
- \frac{1}{6}(\k{}^3+3\k{}^2+2\k{}) \\
\left({2\U{i+3}{n}-27\U{i+2}{n} + 270 \U{i+1}{n} - 490 \U{i}{n} + 270 \U{i-1}{n-\k{}-3} - 27\U{i-2}{n-\k{}-3} + 2\U{i+3}{n-\k{}-3}}\right)/{180\dx^2}  \\
\end{array}$}
&
{ $\mysize \begin{array} {c}  
%{\tiny $\begin{array} {c}  
\frac{49}{864} {\k{}(\k{}+1)(\k{}+2)(\k{}+3)} u^{(0,4)}\frac{\dt^4}{\dx^2}, \\
-\frac{1}{560} u^{(8,0)}{\dx^6}
\end{array}$}
  \\
\hline
%%%%%%%%%%%%%%%%
\end{tabu}}
\caption{Asynchrony-tolerant (AT) schemes for left boundary used in numerical simulations (in \rsec{simulations}). The name of the scheme is represented by the triplet $(d,a,r)$. Two distinguish between two different schemes have the same triplet, we added "a" and "b" to the triplet. Note: minus sign ($-$), if present, has been dropped in leading order terms.}
\label{tab:atschemes-left}
\end{center}
\end{table}
\end{landscape}
    \clearpage% Flush page
}

% table
\afterpage{%
    \clearpage% Flush earlier floats (otherwise order might not be correct)
    \thispagestyle{empty}% empty page style (?)
\begin{landscape}
\begin{table}[h]
\begin{center}
{\tabulinesep=0.5mm
\begin{tabu}{|c|c|c|}
\hline
Scheme & {Scheme at} & Leading order terms \\
$(d,a,r)$ & right boundary & \\
\hline
\hline
%%%%%%%%%%%%%%%%
$(2,1,2)$
%,  uniform delay 
 & 
{ $
\mysize
\begin{array} {c}  
%{\tiny $\begin{array} {c}  
%\frac{\U{i+2}{n-\k{}}-\U{i+1}{n-\k{}}-\U{i}{n}+\U{i-1}{n}}{2\dx^2}
\left({\U{i+2}{n-\k{}}-\U{i+1}{n-\k{}}-\U{i}{n}+\U{i-1}{n}}\right)/{2\dx^2}
\end{array}$}
&
{ $
\mysize
\begin{array} {c}  
%{\tiny $\begin{array} {c}  
\frac{1}{2} \k{} u^{(1,1)}\frac{\dt}{\dx},
\frac{1}{2} u^{(3,0)}{\dx}
\end{array}$}
  \\
\hline
%%%%%%%%%%%%%%%%
$(1,2,2)a$ & %, uniform delay & 
{ $\mysize \begin{array} {c}  
%{\tiny $\begin{array} {c}  
%\frac{\U{i+2}{n-\k{}}-\U{i+1}{n-\k{}}+3\U{i}{n}-3\U{i-1}{n}}{4\dx}
\left({\U{i+2}{n-\k{}}-\U{i+1}{n-\k{}}+3\U{i}{n}-3\U{i-1}{n}}\right)/{4\dx}
\end{array}$}
&
{$\mysize \begin{array} {c}  
%{\tiny $\begin{array} {c}  
\frac{1}{4} {\k{}} u^{(1,1)}\dt,
\frac{5}{12} u^{(3,0)}{\dx^2}
\end{array}$}
  \\
\hline
%%%%%%%%%%%%%%%
$(2,2,2)a$ & %, uniform delay & 
{ $\mysize \begin{array} {c}  
%{\tiny $\begin{array} {c}  
\left({\U{i+3}{n-\k{}}-2\U{i+2}{n-\k{}}+\U{i+1}{n-\k{}}+2\U{i}{n}-4\U{i-1}{n}+2\U{i-2}{n}}\right)/{3\dx^2}
\end{array}$}
&
{ $\mysize \begin{array} {c}  
%{\tiny $\begin{array} {c}  
\frac{1}{3} {\k{}} u^{(2,1)}\dt,
\frac{13}{12} u^{(4,0)}{\dx^2}
\end{array}$}
  \\
\hline
%%%%%%%%%%%%%%%%
$(1,2,2)b$ & %, uniform delay &
{ $\mysize \begin{array} {c}  
%{\tiny $\begin{array} {c}  
\left({(\k{}+1)\U{i+1}{n-\k{}}-\k{}\U{i+1}{n-\k{}-1}-\U{i-1}{n}}\right)/{2\dx}
\end{array}$}
&
{ $\mysize \begin{array} {c}  
%{\tiny $\begin{array} {c}  
\frac{1}{6} u^{(3,0)}{\dx^2}
\end{array}$}
  \\
\hline
%%%%%%%%%%%%%%%%
$(2,2,2)b$ & %, uniform delay &
{ $\mysize \begin{array} {c}  
%{\tiny $\begin{array} {c}  
\left({(\k{}+1)\U{i+1}{n-\k{}}-\k{}\U{i+1}{n-\k{}-1}-2\U{i}{n}+\U{i-1}{n}}\right)/{\dx^2}
\end{array}$}
&
{ $\mysize \begin{array} {c}  
%{\tiny $\begin{array} {c}  
\frac{1}{2} {\k{}(\k{}+1)} u^{(0,2)}\frac{\dt^2}{\dx^2},
\frac{1}{12} u^{(4,0)}{\dx^2}
\end{array}$}
  \\
\hline
%%%%%%%%%%%%%%%%
$(1,4,2)$ & %, uniform delay &
{ $\mysize \begin{array} {c}  
%{\tiny $\begin{array} {c}  
\frac{1}{2}(\k{}^2+3\k{}+2)\left({-\U{i+2}{n-\k{}} + 8 \U{i+1}{n-\k{}} - 8\U{i-1}{n} + \U{i-2}{n}}\right)/{12\dx}  \\
 -(\k{}^2+2\k{})\left({-\U{i+2}{n-\k{}-1} + 8 \U{i+1}{n-\k{}-1} - 8 \U{i-1}{n} + \U{i-2}{n}}\right)/{12\dx} \\
 +\frac{1}{2}(\k{}^2+\k{})\left({-\U{i+2}{n-\k{}-2} + 8 \U{i+1}{n-\k{}-2} - 8 \U{i-1}{n} + \U{i-2}{n}}\right)/{12\dx}
\end{array}$}
&
{ $\mysize \begin{array} {c}  
%{\tiny $\begin{array} {c}  
\frac{1}{30} u^{(5,0)}{\dx^4}
\end{array}$}
  \\
\hline
%%%%%%%%%%%%%%%%
$(2,4,2)$ & %, uniform delay &
{ $\mysize \begin{array} {c}  
%{\tiny $\begin{array} {c}  
\frac{1}{2}(\k{}^2+3\k{}+2) \left({-\U{i+2}{n-\k{}} + 16 \U{i+1}{n-\k{}} - 30 \U{i}{n} + 16\U{i-1}{n} - \U{i-2}{n}}\right)/{12\dx^2}  \\
 -(\k{}^2+2\k{}) \left({-\U{i+2}{n-\k{}-1} + 16 \U{i+1}{n-\k{}-1} - 30 \U{i}{n} + 16\U{i-1}{n} - \U{i-2}{n}}\right)/{12\dx^2}  \\
 +\frac{1}{2}(\k{}^2+\k{}) \left({-\U{i+2}{n-\k{}-2} + 16 \U{i+1}{n-\k{}-2} - 30 \U{i}{n} + 16\U{i-1}{n} - \U{i-2}{n}}\right)/{12\dx^2} 
\end{array}$}
&
{ $\mysize \begin{array} {c}  
%{\tiny $\begin{array} {c}  
\frac{5}{24} {\k{}(\k{}+1)(\k{}+2)} u^{(0,3)}\frac{\dt^3}{\dx^2}, \\
\frac{1}{90} u^{(6,0)}{\dx^4}
\end{array}$}
  \\
\hline
%%%%%%%%%%%%%%%%
$(2,6,2)$ & %, uniform delay &
{ $\mysize \begin{array} {c}  
%{\tiny $\begin{array} {c}  
\frac{1}{6}(\k{}^3+6\k{}^2+11\k{}+6) \\
\left({2\U{i+3}{n-\k{}}-27\U{i+2}{n-\k{}} + 270 \U{i+1}{n-\k{}} - 490 \U{i}{n} + 270 \U{i-1}{n} - 27\U{i-2}{n} + 2\U{i-3}{n}}\right)/{180\dx^2}  \\
- \frac{1}{2}(\k{}^3+5\k{}^2+6\k{}) \\
\left({2\U{i+3}{n-\k{}-1}-27\U{i+2}{n-\k{}-1} + 270 \U{i+1}{n-\k{}-1} - 490 \U{i}{n} + 270 \U{i-1}{n} - 27\U{i-2}{n} + 2\U{i-3}{n}}\right)/{180\dx^2}  \\
+ \frac{1}{2}(\k{}^3+4\k{}^2+3\k{}) \\
\left({2\U{i+3}{n-\k{}-2}-27\U{i+2}{n-\k{}-2} + 270 \U{i+1}{n-\k{}-2} - 490 \U{i}{n} + 270 \U{i-1}{n} - 27\U{i-2}{n} + 2\U{i-3}{n}}\right)/{180\dx^2}  \\
- \frac{1}{6}(\k{}^3+3\k{}^2+2\k{}) \\
\left({2\U{i+3}{n-\k{}-3}-27\U{i+2}{n-\k{}-3} + 270 \U{i+1}{n-\k{}-3} - 490 \U{i}{n} + 270 \U{i-1}{n} - 27\U{i-2}{n} + 2\U{i-3}{n}}\right)/{180\dx^2}  \\
\end{array}$}
&
{ $\mysize \begin{array} {c}  
%{\tiny $\begin{array} {c}  
\frac{49}{864} {\k{}(\k{}+1)(\k{}+2)(\k{}+3)} u^{(0,4)}\frac{\dt^4}{\dx^2}, \\
-\frac{1}{560} u^{(8,0)}{\dx^6}
\end{array}$}
  \\
\hline
%%%%%%%%%%%%%%%%
\end{tabu}}
\caption{Asynchrony-tolerant (AT) schemes for right boundary used in numerical simulations (in \rsec{simulations}). The name of the scheme is represented by the triplet $(d,a,r)$. Two distinguish between two different schemes have the same triplet, we added "a" and "b" to the triplet. Note: minus sign ($-$), if present, has been dropped in leading order terms.}
\label{tab:atschemes-right}
\end{center}
\end{table}
\end{landscape}
    \clearpage% Flush page
}

% simulate delays
In numerical simulations, we use a random number generator to simulate communication delays ($\k{j}$) at PE boundaries.
This provides a complete control over the statistics of the delays, thus, allowing us to compare the results against the theoretical predictions in different parameter regimes. 
At each time advancement, the delay at a buffer point is computed from a random number drawn with a given initial seed from a uniform distribution in the interval $[0,1]$.
This interval is divided into $L$ bins according to the probabilities $\{\prob{0}{j}, \prob{1}{j}, \dots, \prob{L-1}{j}\}$
corresponding to delays $\k{j}=0, 1, 2, \dots, L-1$, respectively.
When a random number is drawn, it is matched with the corresponding bin which determines the delay.
As we use i.i.d.\ random sequences at different PE boundaries in the simulations, there is no dependence on the location and hence, we drop the subscript $j$ in probabilities for simplicity and write $\{p_{0}, p_{1}, \dots, p_{L-1}\}$.
As an example, if we choose $L=3$ and the set $\{p_{0}, p_{1},
p_{2}\}=\{0.6,0.3,0.1\}$, then the probability of having $\k{}=0$, $\k{}=1$
and $\k{}=2$ is 
$0.6$, $0.3$ and $0.1$, respectively.
In the case of schemes which use a uniform delay in their stencil (like in \req{cd4-at} of Example 3), a single random number is drawn at each PE boundary to obtain the uniform delay at all the buffer points at that PE boundary. 

% stability
%In this work our objective is to demonstrate the robustness of AT 
%schemes in regard to accuracy. 
%The details of the effect of the schemes on numerical stability will be a topic
%of interest elsewhere.
%For the simulations in \rtab{cases}, the stability for combination of
%schemes was verified through numerical experiments.
The error is computed by comparing the numerical solution 
against the analytical solution.
With periodic boundary conditions and an initial condition given in \req{IC}, the analytical solution (denoted by subscript $a$) for the linear advection-diffusion equation is
\be
u_a(x,t) = \sum_{\kappa}e^{-\alpha\kappa^2 t}A(\kappa)\sin(\kappa x + \phik -ct).
\label{eq:ana_sol}
\ee
For the heat equation, the analytical solution is given by the above expression with $c=0$.
In the case of nonlinear Burgers equation, the error is evaluated against the solution from a highly resolved simulation. 
The error at a point $i$ and time level $n$ is computed as $\E{i}{n}=\U{i}{n}-u_a(x_i,t_n)$. The overall error in the domain is obtained using the different averages presented in \rsec{error}.

%\end{comment}
\subsection{Results}\label{sec:results}
\subsubsection{Linear equations}
\begin{figure}[h]
  \centering
  \subfigure{\includegraphics[width=0.49\textwidth]{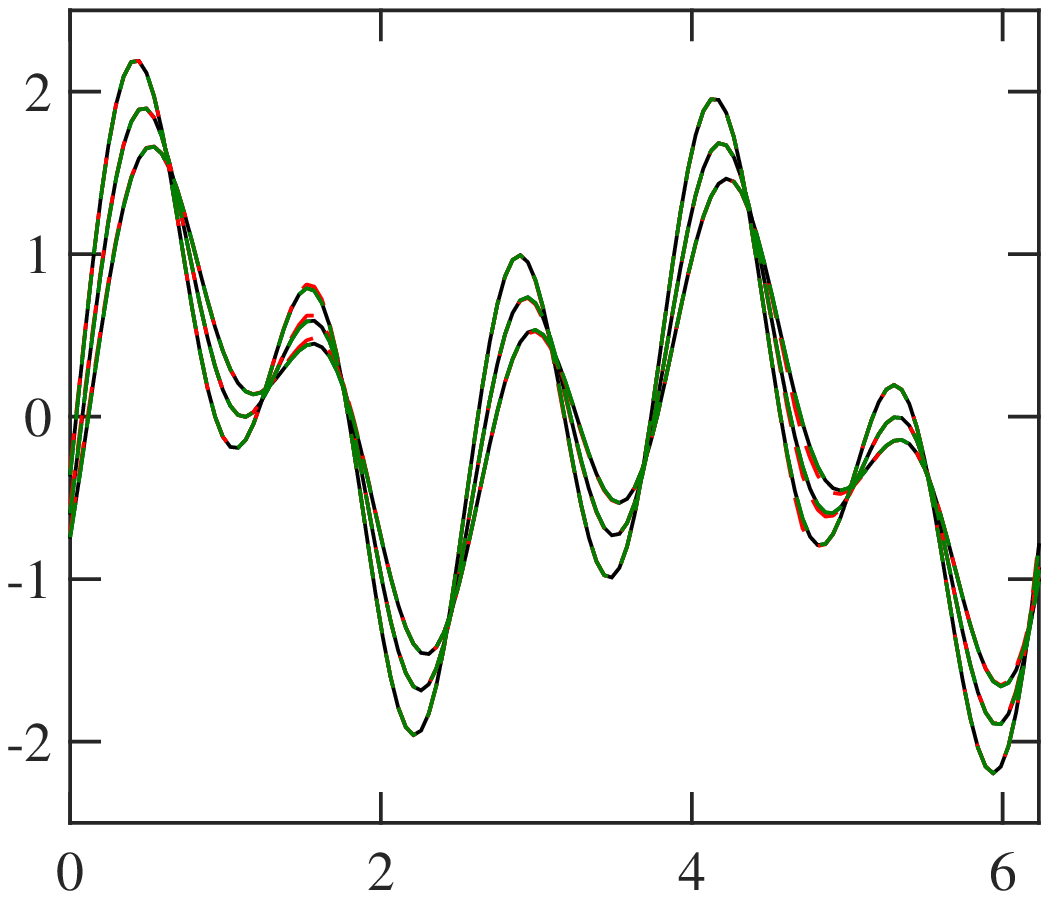}}
  \subfigure{\includegraphics[width=0.49\textwidth]{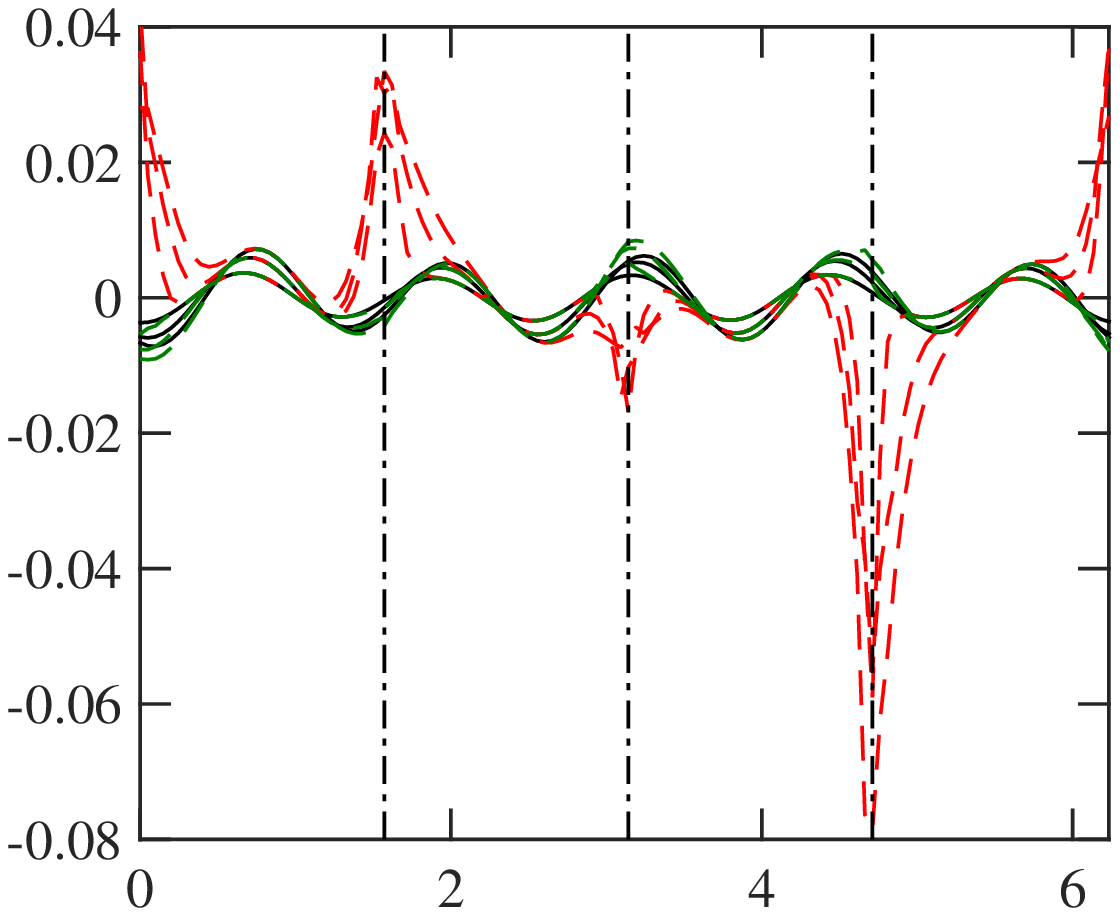}}
   \begin{picture}(0,0)
      \put(-260,0){$x$}
      \put(-85,0){$x$}
      \put(-340,65){\rotatebox{90} {$u$}}
      \put(-178,55){\rotatebox{90} {$u-u_a$}}
      \put(-310,60){\vector(1,2){15}}
      \put(-120,80){\vector(1,1){20}}
      \put(-295,95){$t$}
      \put(-98,105){$t$}
      \put(-215,115){(a)}
      \put(-40,115){(b)}
   \end{picture}
  \caption{Typical time evolution of the numerical solution of the
  advection-diffusion equation 
  using synchronous (solid black lines), asynchronous standard (dashed red lines) and \ats schemes (green lines).
  (a) The velocity field. (b) Error $\E{i}{n}=\U{i}{n} - u_a(x_i,t_n)$.
  Vertical dash-dotted lines correspond to PE boundaries.
  Simulation parameters: $N=128$, $P=4$, $L=3$, with
  $\{p_0,p_1,p_2\}=\{0.5,0.3,0.2\}$ for the asynchronous computations.}
  \label{fig:at-sol1}
\end{figure}

\rfig{at-sol1} shows results from simulations with three different schemes:
synchronous (solid black lines), asynchronous-standard (dashed red lines) and
AT (green lines) schemes.
Note that by asynchronous-standard schemes we mean standard (synchronous)
schemes used in an asynchronous fashion.
The governing equation and schemes are those corresponding to Case 1  in
\rtab{cases}. The simulation parameters are $N=128$, $\kappa=\{2,3,5\}$ (the
vector $\kappa$ here contains the wavenumbers used in the initial condition
defined by \req{ana_sol}), $P=4$, and three allowable time levels for
asynchronous computations according to $\{p_0,p_1,p_2\}=\{0.5,0.3,0.2\}$. 
In part (a) of the figure, we show the time evolution of function $u$.
We observe that the initial condition which is a combination of sine waves is convected with a wave speed $c$ and simultaneously damped due to diffusive action, as expected.
%No distinct differences are noticed in the solution from the three simulations.
%However, in part (b), differences are evident in the evolution of error.
To highlight the differences between these cases, we show the evolution of the error in part (b) of the figure.
The error in the case of asynchronous standard schemes is an order of magnitude greater near the PE boundaries (indicated by vertical dash-dotted lines).
As discussed in \cite{DA2014}, this is due to the asynchrony in the data available at buffer points. This error, which is initially localized near PE boundaries, propagates into the interior with time.
In the case of \ats schemes, which are designed to mitigate the affect of asynchrony, the error at PE boundaries is of similar magnitude as the synchronous schemes.

\begin{figure}[h]
  \centering
  \subfigure{\includegraphics[width=0.49\textwidth]{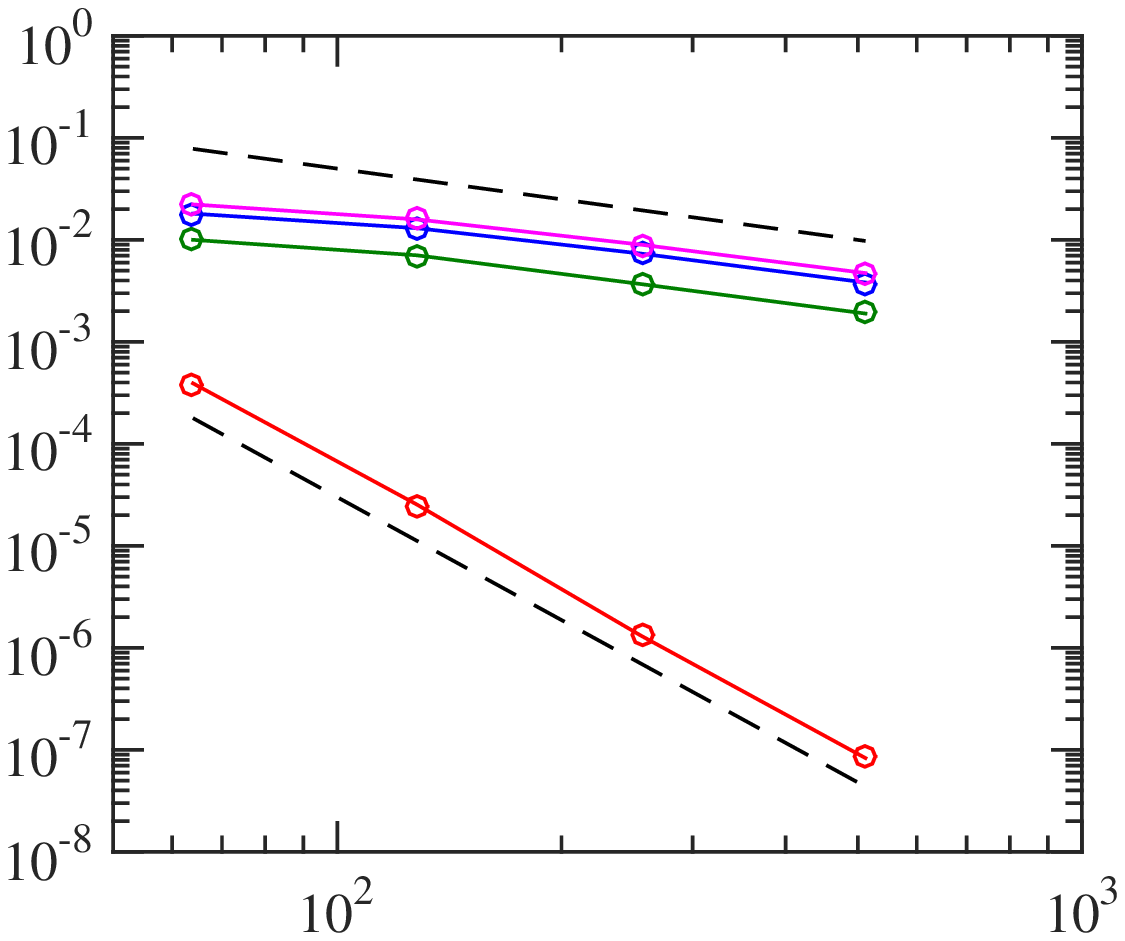}}
   \begin{picture}(0,0)
      \put(-85,0){$N$}
      \put(-180,65){\rotatebox{90} {$\ave{E}$}}
      \put(-85,105){$-1$}
      \put(-95,30){$-4$}
      \put(-85,0){$N$}
   \end{picture}
  \caption{ Convergence plot of the average overall error with increasing grid
  resolution. Results are obtained from the simulations of advection-diffusion
  equation with fourth order standard central difference schemes. Different
  lines correspond to varying degree of asynchrony introduced in the
  simulations: $p_0=1.0$ (red), $p_0=0.7$ (green), $p_0=0.3$ (blue) and
  $p_0=0.0$ (magenta). Dashed lines with a slope of $-1$ and $-4$ are shown for
  reference.}
  \label{fig:at-ord4-async}
\end{figure}

To verify the formal order of accuracy of AT schemes and the effect of simulation parameters ($N$, $P$, $\kmom{}$, etc.) on the overall error, we now proceed to the statistical description of the error.
An example of the effect of asynchrony on the overall error for standard
central difference schemes (Case 4 in \rtab{cases}) is shown in \rfig{at-ord4-async}. 
%The results in
%this figure are obtained from the simulations of Case(4) in \rtab{cases}. 
For the fourth-order scheme used in these simulations, the error for $p_0=1.0$ decreases with a slope of $-4$, as expected in synchronous computing.
In the presence of asynchrony ($p_0<1.0$) the slope reduces to $-1$, depicting
a first order accurate solution \cite{DA2014}. Also, the absolute error for a given grid
resolution increases when asynchrony is increased ($p_0$ is reduced).
This drastic decrease in accuracy is mitigated when 
AT schemes are used, as we show next.

\begin{figure}[h]
  \centering
  \subfigure{\includegraphics[width=0.49\textwidth]{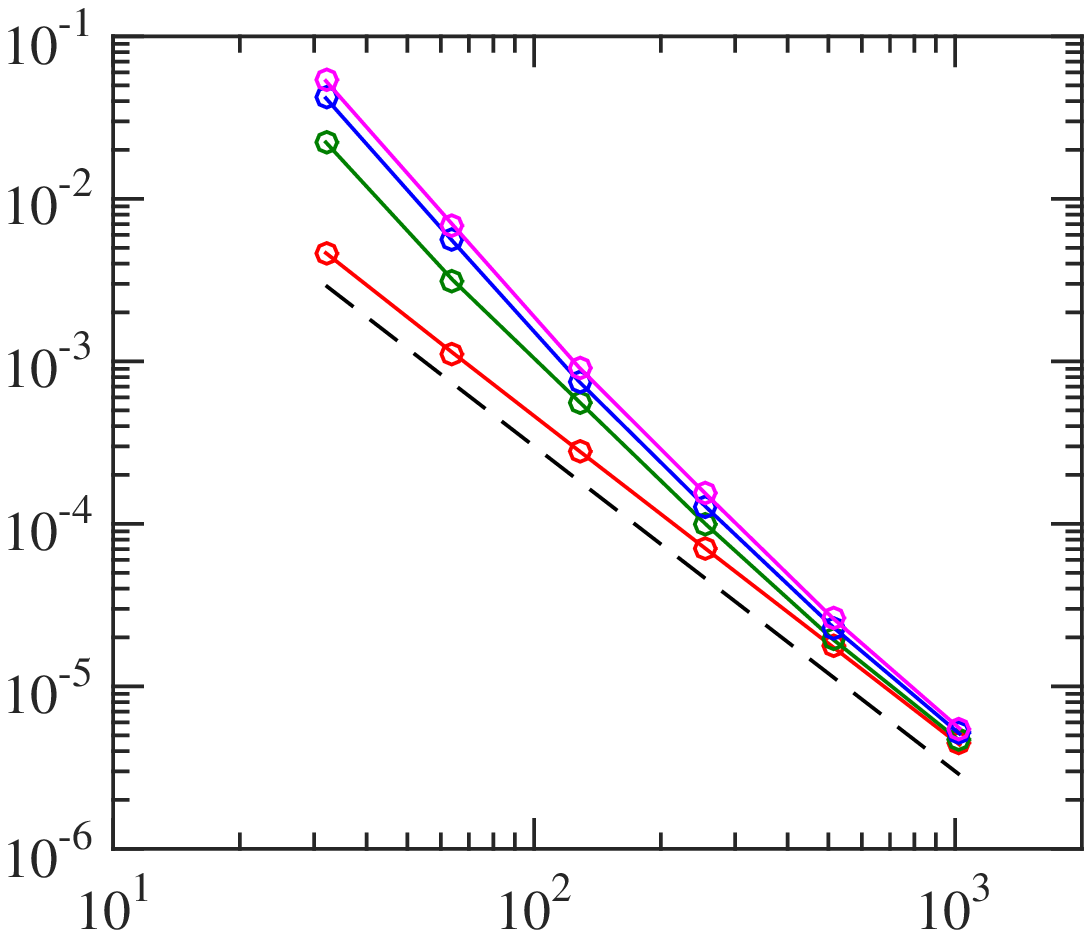}}
  \subfigure{\includegraphics[width=0.49\textwidth]{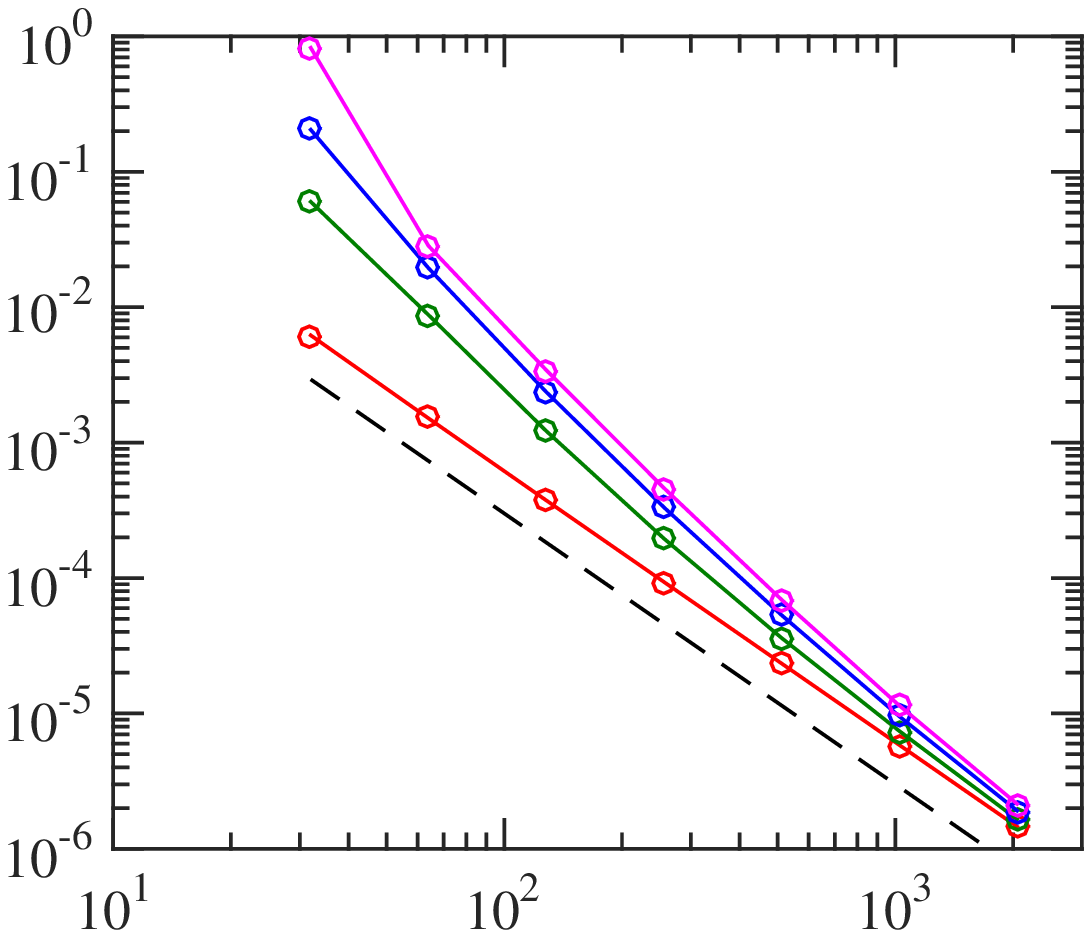}}
  \subfigure{\includegraphics[width=0.49\textwidth]{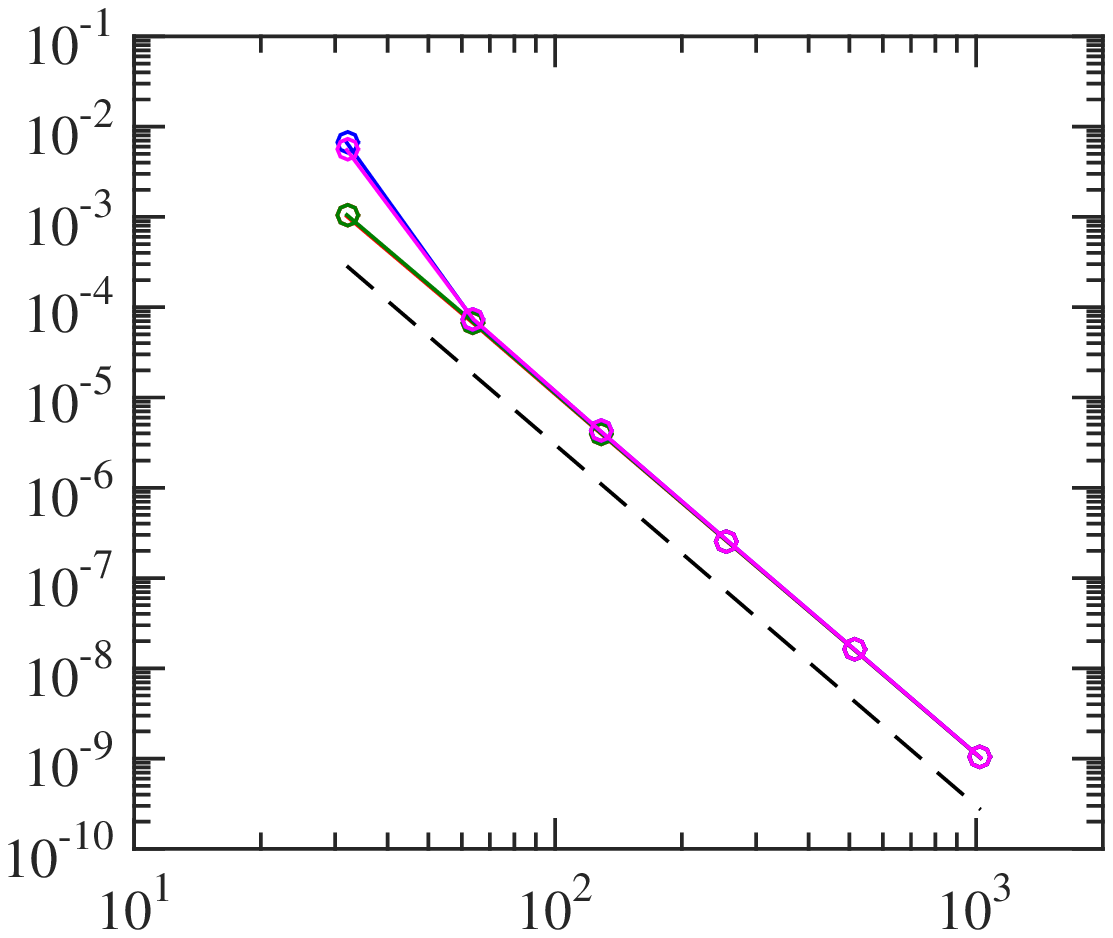}}
  \subfigure{\includegraphics[width=0.49\textwidth]{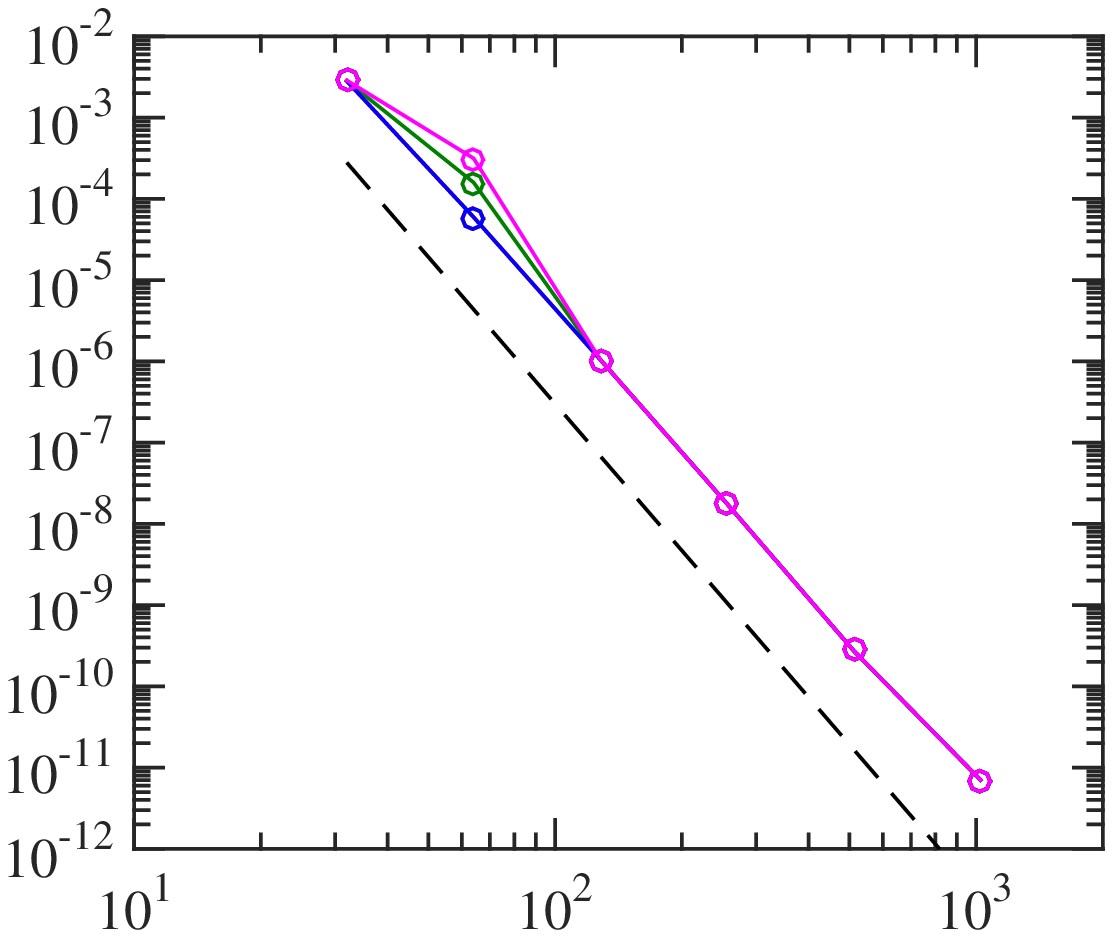}}
   \begin{picture}(0,0)
      \put(-260,0){$N$}
      \put(-85,0){$N$}
      \put(-260,140){$N$}
      \put(-85,140){$N$}
      \put(-350,62){\rotatebox{90} {$\ave{E}$}}
      \put(-180,62){\rotatebox{90} {$\ave{E}$}}
      \put(-350,202){\rotatebox{90} {$\ave{E}$}}
      \put(-180,202){\rotatebox{90} {$\ave{E}$}}
      \put(-218,252){(a)}
      \put(-40,252){(b)}
      \put(-218,112){(c)}
      \put(-40,112){(d)}
      \put(-268,190){$-2$}
      \put(-278,55){$-4$}
      \put(-95,180){$-2$}
      \put(-100,55){$-6$}
   \end{picture}
  \caption{ Convergence plot of the average overall error for Cases 2, 3, 4 and
  5 listed in \rtab{cases}, simulated with \ats schemes at
  communication delayed buffer points. Different lines in each graph correspond
  to a varying degree of asynchrony introduced in the simulations: $p_0=1.0$
  (red), $p_0=0.7$ (green), $p_0=0.3$ (blue) and $p_0=0.0$ (magenta). Dashed
  lines with constant slope (value shown adjacent to line) shown for
  reference.}
  \label{fig:at-prob-all}
\end{figure}

\rfig{at-prob-all} shows the effect on asynchrony in different configurations when AT schemes are used. The parameters used in these numerical experiments are: $\kappa=\{3,4,5\}$, $A(\kappa)=\{2.0,0.5,1.5\}$, $P=16$, $L=3$.
Different colors in the graphs represent results from different sets of $p_k$ ($k=0,1,2$) and their values are given in the caption of the figure.
%, $\{p_0,p_1,p_2\}=\left[ \{1,0,0\},\{0.7,0.3,0.0\},\{0.5,0.3,0.2\},\{0.4,0.4,0.2\} \right]$.
Part (a) shows results from Case 2 in \rtab{cases}. A second-order central
difference scheme for synchronous computations and a first-order asymmetric
stencil AT scheme for asynchronous computations are used.
The error in absence of delays (red line) decreases
with a slope of $-2$ as expected.
We also observe that, asymptotically,
this is also the case for the asynchronous cases ($p_0<1$).
%When asynchrony is present, the error decreases with increase in grid points
%and asymptotes to a slope of $-2$, representing an overall second order
%accurate solution.
The reason for second-order accuracy even with a first-order AT scheme can be
explained with the strong scaling argument presented in \req{aveE_2}. 
The effect of an 
increase in the amount of asynchrony, is seen to increase the magnitude of 
error leaving the asymptotic rate of convergence unchanged. 

% part (b)
In part (b), we show results for Case 3 in \rtab{cases}.
A second-order accurate asymmetric stencil AT
scheme is used for asynchronous computations. In this case too, we see an
asymptotic convergence rate of order 2. Also, the magnitude of error increases
with the amount of asynchrony.
Note that in both parts (a) and (b), the AT schemes are constructed by
expanding the stencil in space to improve the accuracy in the presence of
asynchrony.
These schemes do not reduce to or have the same form as the central difference
schemes when $\k{}=0$. 

In parts (c) and (d) of \rfig{at-prob-all}, results are shown 
for AT schemes derived by
expanding the stencil in time (instead of space) to maintain accuracy in the
presence of
asynchrony. These are cases 4 and 5 in \rtab{cases}.
These schemes have symmetric stencils and coefficients, and they
reduce to central difference schemes when $\k{}=0$. 
%The configurations for the
%two graphs are given in Cases (4) and (5) of \rtab{cases}.
As expected from the theory, the error in part (c) converges with an accuracy of
order 4. The effect of asynchrony is hardly noticeable at higher resolutions.
This can be attributed to the fact that the AT schemes, in this case, reduce to
the synchronous central schemes in absence of asynchrony and result in a
homogeneous synchronous truncation error terms across the domain. A similar
observation is also found in part (d), which uses a sixth order accurate scheme
for space derivative. 

\begin{figure}[h]
  \centering
  \subfigure{\includegraphics[width=0.49\textwidth]{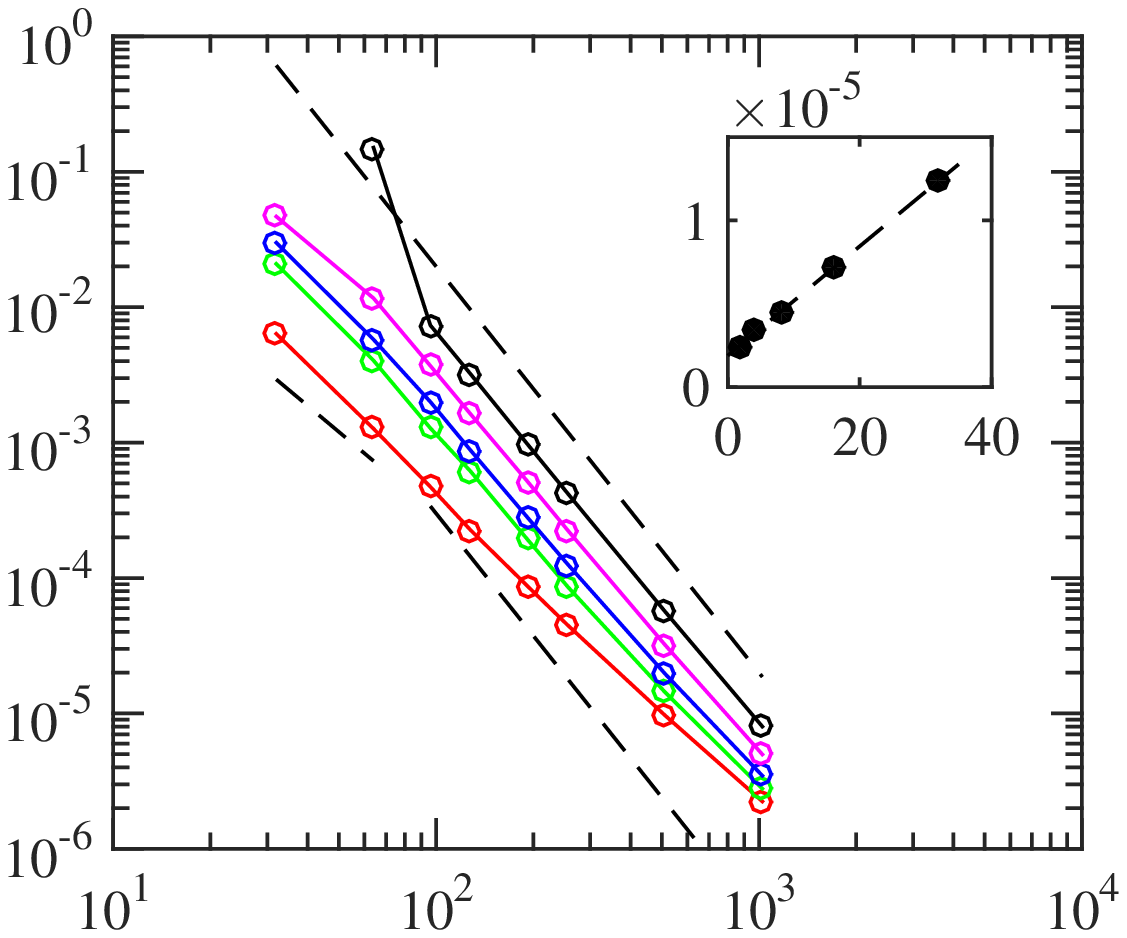}}
  \subfigure{\includegraphics[width=0.49\textwidth]{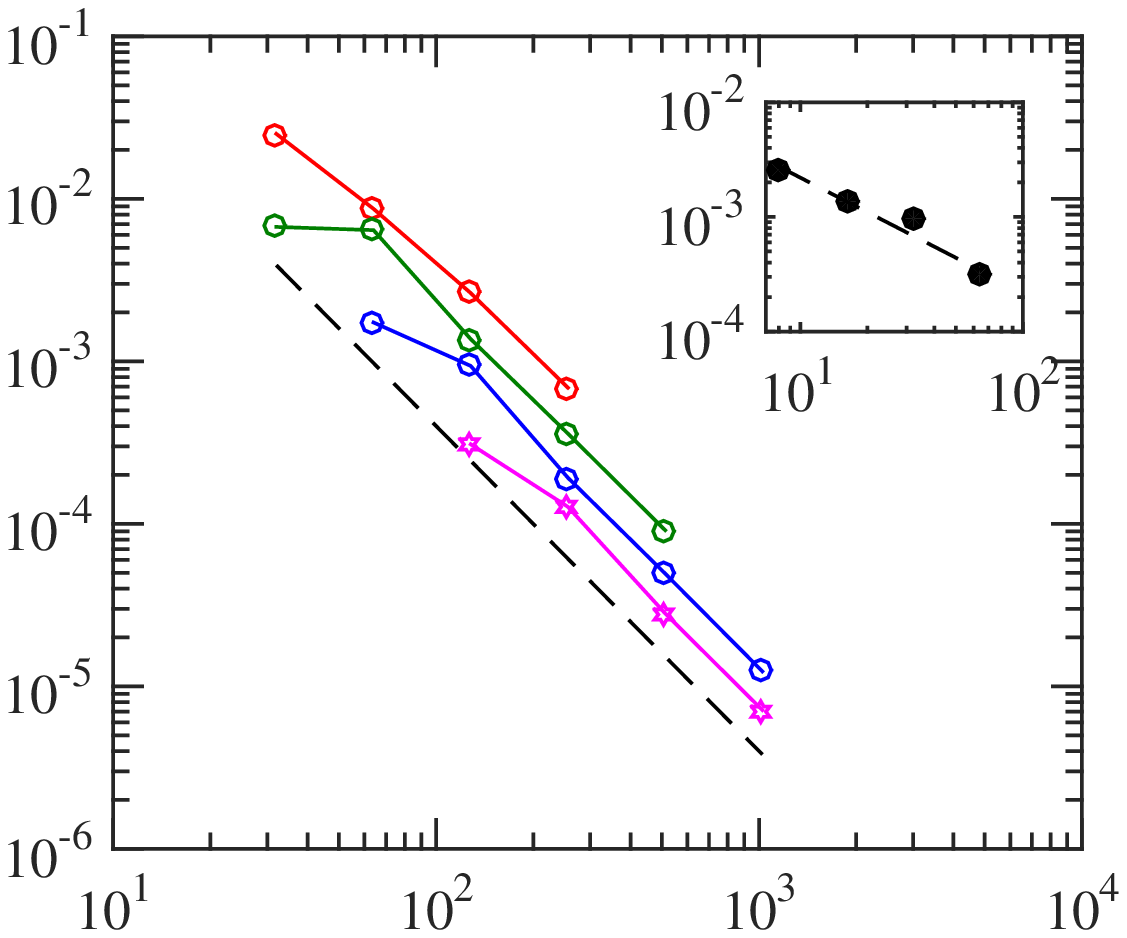}}
   \begin{picture}(0,0)
      \put(-260,0){$N$}
      \put(-85,0){$N$}
      \put(-350,60){\rotatebox{90} {$\ave{E}$}}
      \put(-178,65){\rotatebox{90} {$\ave{E}$}}
      \put(-220,60){\small $P$}
      \put(-53,70){\small $N/P$}
      \put(-315,105){(a)}
      \put(-145,105){(b)}
      \put(-308,63){$-2$}
      \put(-280,35){$-3$}
      \put(-270,85){$-3$}
      \put(-110,50){$-2$}
   \end{picture}
  \caption{Effect of number of PEs on the average error for the Case 1 with 
$L=3$, $p_k=\{0.2,0.5,0.3\}$ and $r_\alpha = 0.1$.
  (a)
  Strong scaling: cases with constant $P$. Different lines correspond to $P =
  2$ (red), $4$ (green), $8$ (blue), $16$ (magenta), $32$ (black). Inset: plot
  of average error with $P$ at $N=512$. (b) Weak scaling: cases with constant
  $P/N$. Different lines correspond to $P/N = 1/64$ (magenta), $1/32$ (blue),
  $1/16$ (green), $1/16$ (red). Inset: plot of average error with $N/P$ at
  $N=128$. Dashed lines with constant slope (value shown adjacent to line) are
  included for reference.}
  \label{fig:at-ord2-procs}
\end{figure}

% strong and weak scaling
Let us now recall \req{aveE_scaling_gen}, which describes the scaling of the average error $(\ave{E})$ with $P$, $N$ and $\k{}$:
\bea
\ave{E} & \sim & \frac{P}{N} \dx^a \sum_{m=1}^{\mathcal{T}}\gamma_m\kmom{m} \nonumber \\
& \sim & {P}\dx^{a+1} \sum_{m=1}^{\mathcal{T}}\gamma_m\kmom{m}
\label{eq:aveE_scaling_gen2}
\eea
Note that this scaling holds only when the error due to asynchrony
dominates the overall error.
Otherwise, the error in the leading order may have a synchronous component 
and may show a different dependence on simulation
parameters.

In \rfig{at-ord2-procs} 
%supports the proposed linear scaling of error with $P$ using numerical
we show numerical data from Case 1.
According to \req{aveE_scaling_gen2},
the order of accuracy is one more than the order of the AT scheme
when $P$ is fixed.
%under strong scaling, that is when the problem size, $N$, is kept constant 
%while $P$ is increased.
Part (a) of the figure shows the convergence of the error for different $P$.
For low $P$, the leading order terms of the error contain both synchronous and
asynchronous
contributions, and thus shows a convergence slope between $-3$ and $-2$.
However, when $P$ increases, the error due to asynchrony dominates and shows a
convergence of $-3$, as predicted by the theory.
The linear scaling of the error with
$P$ is verified in the inset of part (a). 
Results for weak scaling, that is when both $P$ and $N$ in increase
such that $P/N$ is kept constant, are shown in part (b).
As expected, the error for different $P/N$ asymptotically converges to second
order accuracy. In the inset of the figure, an inverse dependence of the error on
$N/P$ is observed as predicted also by the theory. 

\begin{figure}[h]
  \centering
  \subfigure{\includegraphics[width=0.49\textwidth]{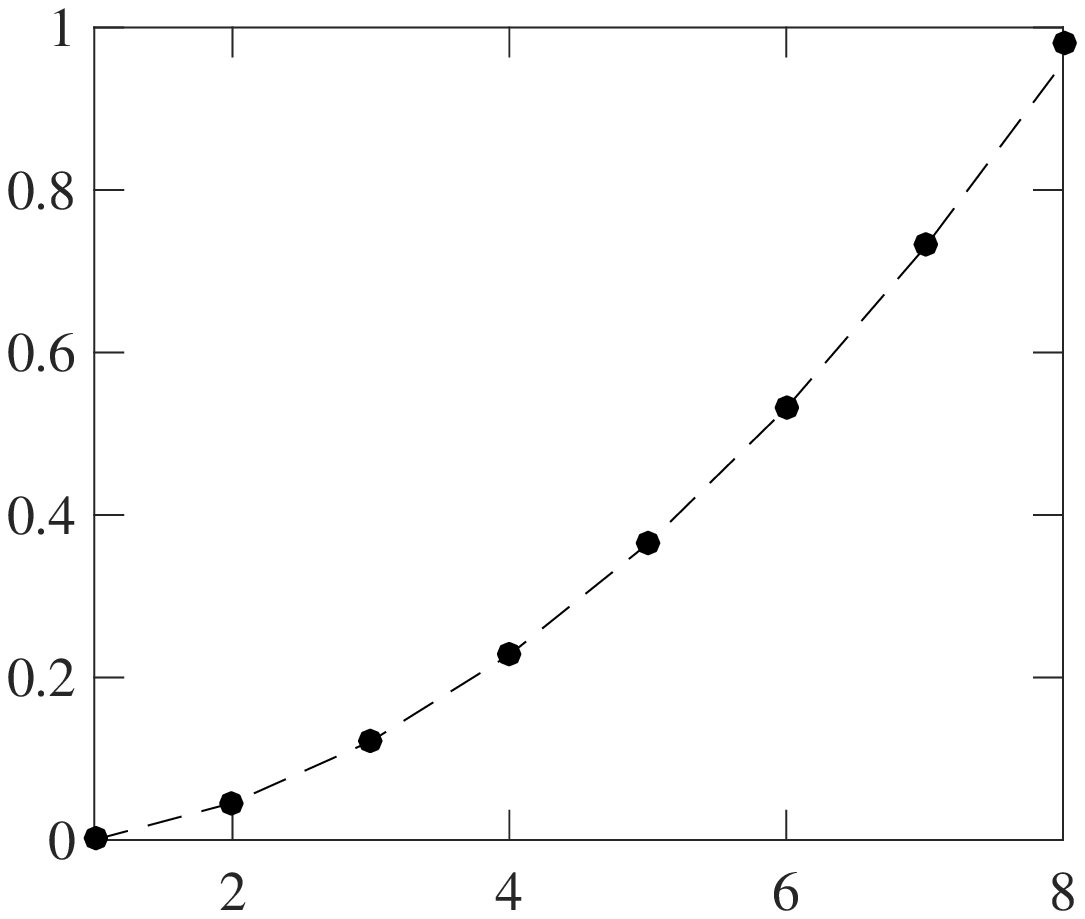}}
  \subfigure{\includegraphics[width=0.49\textwidth]{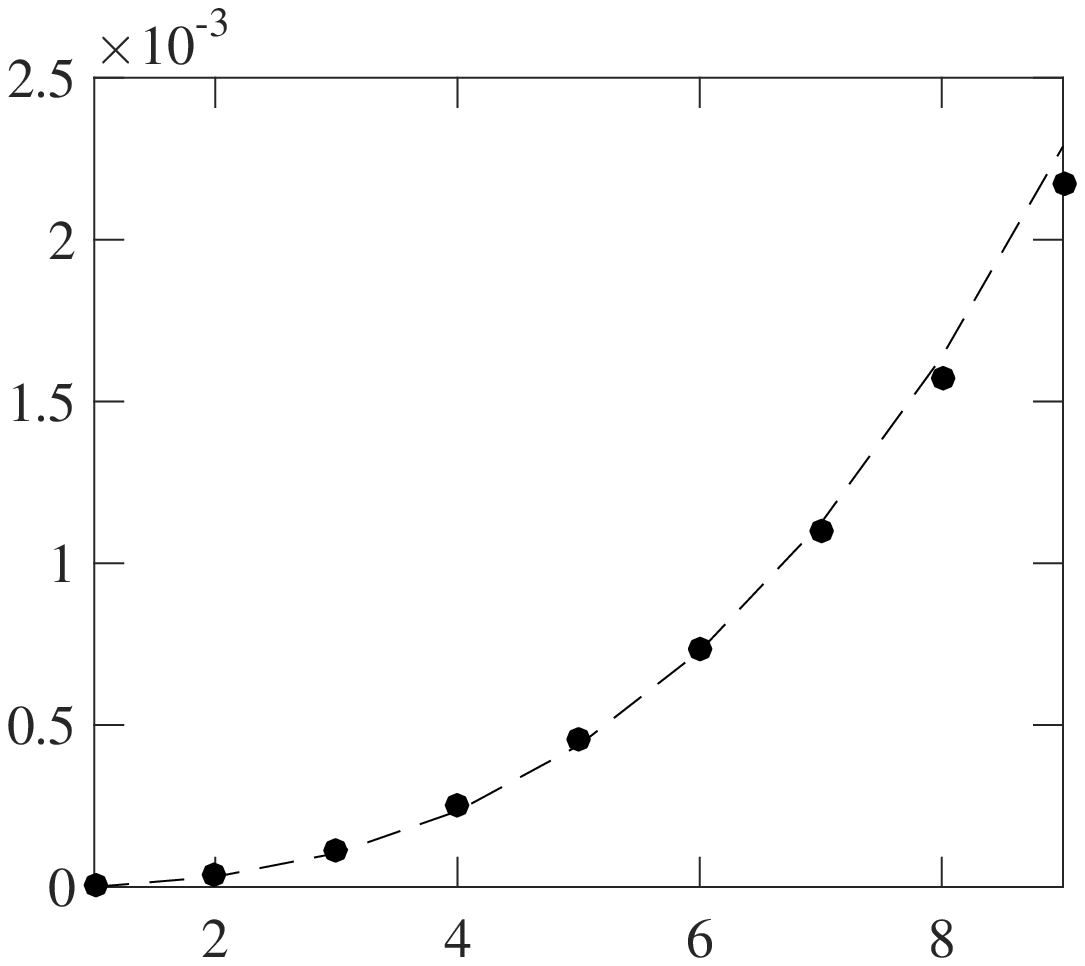}}
   \begin{picture}(0,0)
      \put(-260,-4){$L$}
      \put(-85,-4){$L$}
      \put(-348,30){\rotatebox{90} {$(\ave{E}-\ave{E}_s)/\ave{E}_s$}}
      \put(-177,30){\rotatebox{90} {$(\ave{E}-\ave{E}_s)/\ave{E}_s$}}
      \put(-213,110){(a)}
      \put(-40,110){(b)}
   \end{picture}
  \caption{Scaling of normalized average error ($(\ave{E}-\ave{E}_s)/\ave{E}_s$) with moments of the
  delay ($\k{}$). In parts (a) and (b), circles are obtained from simulation of
  Cases 3 and 4, respectively, with parameters $N=512$ and $P=16$. The
  dashed curves are polynomials in \req{aveE_scaling_gen_L} obtained from
  theory.}
% \textcolor{blue}{[DD] Can we normalize the error so that the
%  order of magnitude of the y axis is not so small, e.g.
%  $(\ave{E}-\ave{E}_s)/\ave{E}_s$? Equations and coefficients should not change.}} 
  \label{fig:at-kmom}
\end{figure}

% moments
Unlike standard synchronous schemes when asynchrony is allowed,
for which the error due to asynchrony 
depends only on the average of delays ($\kmom{}$) \cite{DA2014}, the error
for \ats schemes can also depend on higher order moments of $\k{}$, as
shown in \rsec{error}. 
This dependence is indeed confirmed in \rfig{at-kmom}. Results
in parts (a) and (b) of the figure are for Cases 3 and 4, respectively
and can be understood as follows.
For these schemes, which use two and three delayed time levels
in the stencil
($\mathcal{T}=2$ and $3$), respectively, the average error scales as:
\bea
\ave{E} & \sim & \left( \kmom{2} + \kmom{} \right) \text{\hspace{2.3cm}for  Case 3}  \nonumber \\
 & \sim & \left(\kmom{3}+3 \kmom{2} +2 \kmom{} \right) \text{\hspace{1.15cm}for Case 4} 
\label{eq:aveE_scaling_gen_kmom}
\eea
To simplify the dependence of error to a single variable, the probability of
occurence of a level $k$ for a given $L$ is chosen as $p_k=1/L$.
For example, if $L=3$, then $\{p_0,p_1,p_2\}=\{1/3,1/3,1/3\}$. This reduces
the scaling of the average error to
\bea
\ave{E} & \sim & \left( L^2 - 1  \right) \text{\hspace{2.5cm} for Case 3}  \nonumber \\
 & \sim & \left(L^3 + 2 L^3 - L - 2 \right) \text{\hspace{0.75cm} for Case 4}. 
\label{eq:aveE_scaling_gen_L}
\eea
As mentioned earlier, this scaling is only valid for the asynchrony component of
the error or for the total error when the former is dominant to leading order. 
Thus, to compare to \req{aveE_scaling_gen_L}, we compute the total error
and subtract it from a completely synchronous, but otherwise
identical, simulation.
%To remove the affects of the synchronous
%component of the error we plot, from numerical experiments, 
%the normalized average error
%with respect to the average error in synchronous simulation given by
%$(\ave{E}-\ave{E}_s)/\ave{E}_s$. 
In \rfig{at-kmom} we show the thus obtained asynchronous part of the 
error, that is $(\ave{E}-\ave{E}_s)/\ave{E}_s$. 
The
dashed curves in the graphs correspond to \req{aveE_scaling_gen_L}.
There is very good agreement between the theoretical prediction and the
data from simulations.

\subsubsection{Nonlinear equations}
\begin{figure}[h]
  \centering
  \subfigure{\includegraphics[width=0.49\textwidth]{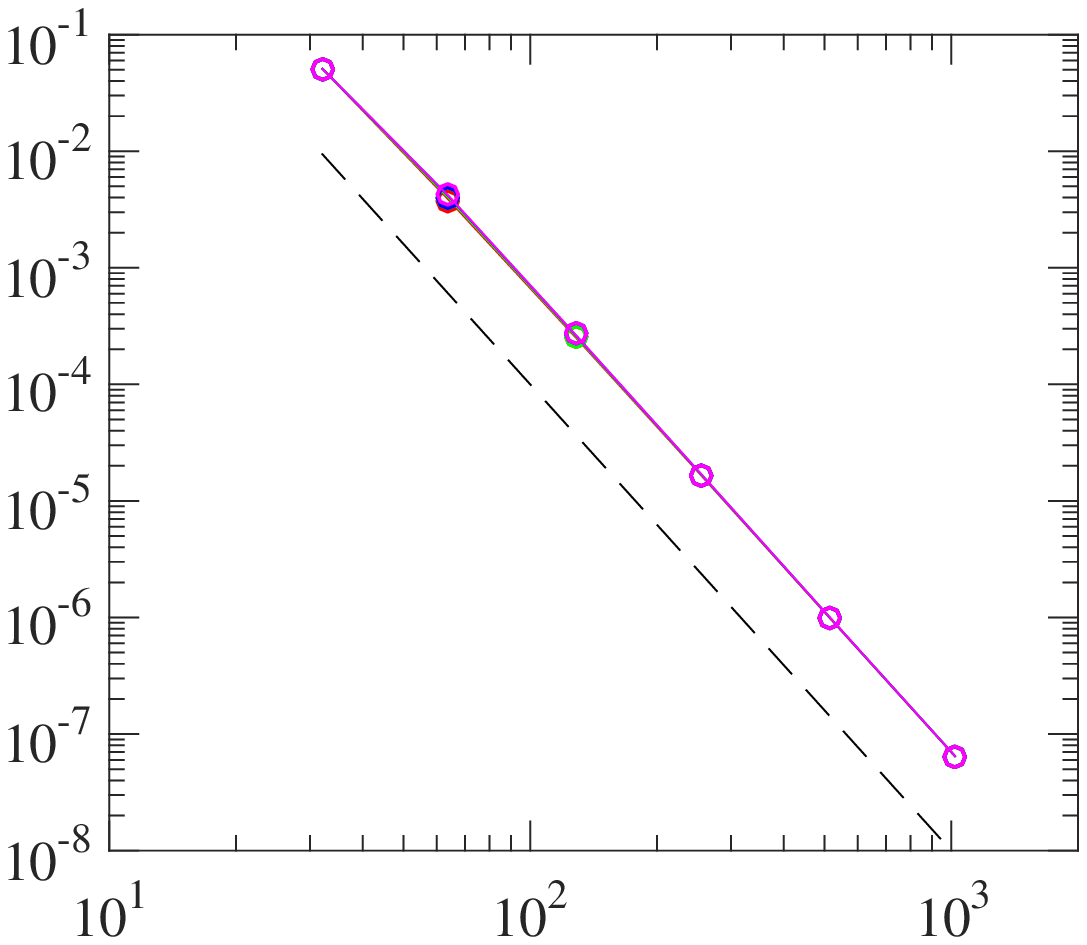}}
   \begin{picture}(0,0)
      \put(-90,-5){$N$}
      \put(-95,55){$-4$}
      \put(-180,60){\rotatebox{90} {$\ave{E}$}}
   \end{picture}
  \caption{ Convergence plot of the average overall error with increasing grid
  resolution. Results are from simulations of the nonlinear viscous
  Burgers' equation (Case 6 in \rtab{cases}).
  Different lines correspond to varying degree of asynchrony introduced in the
  simulations: $p_0=1.0$ (red), $p_0=0.7$ (green), $p_0=0.4$ (blue) and
  $p_0=0.3$ (magenta). Dashed line is a reference power law with slope $-4$.}
  \label{fig:at-ord4-nl}
\end{figure}

In the above numerical experiments, we have verified the performance of \ats
schemes for linear equations.
However, a number of natural and engineering systems are governed by highly
nonlinear processes like fluid turbulence phenomena.
The viscous Burgers' equation is often used as a proxy 
to understand these nonlinear effects in fluid flows
with negligible pressure effects. Thus, we use this equation 
to assess \ats schemes in a more realistic setup.
\rfig{at-ord4-nl} shows the convergence of the error for
the fourth-order schemes described in \rtab{cases} as Case 6.
Clearly, even with an increase in the degree of asynchrony the error converges
with fourth-order accuracy.

\begin{figure}[h]
  \centering
  \subfigure{\includegraphics[width=0.49\textwidth]{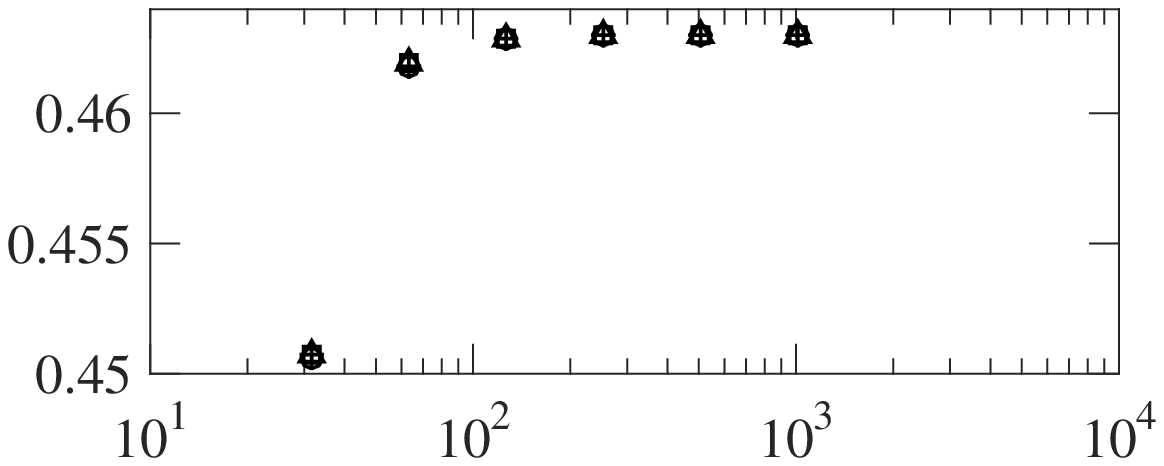}}
  \subfigure{\includegraphics[width=0.49\textwidth]{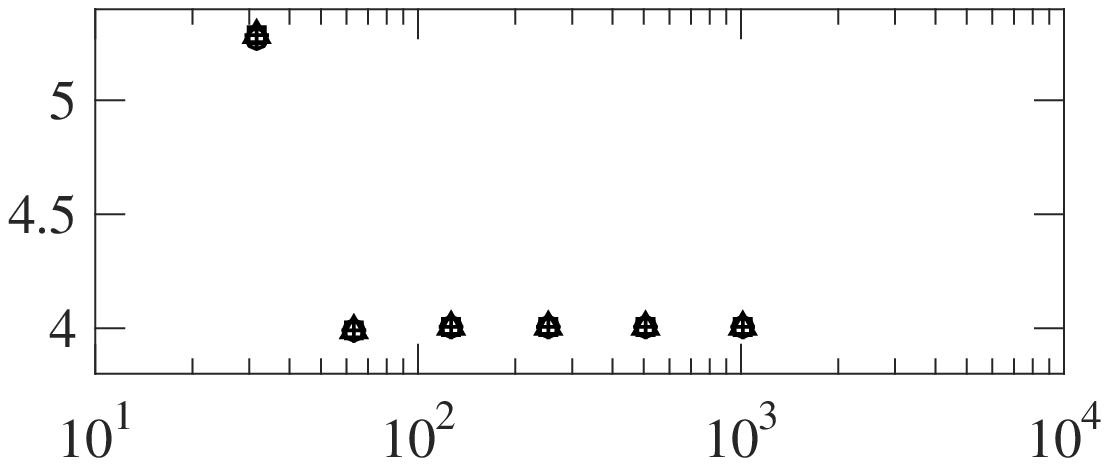}}
  \subfigure{\includegraphics[width=0.49\textwidth]{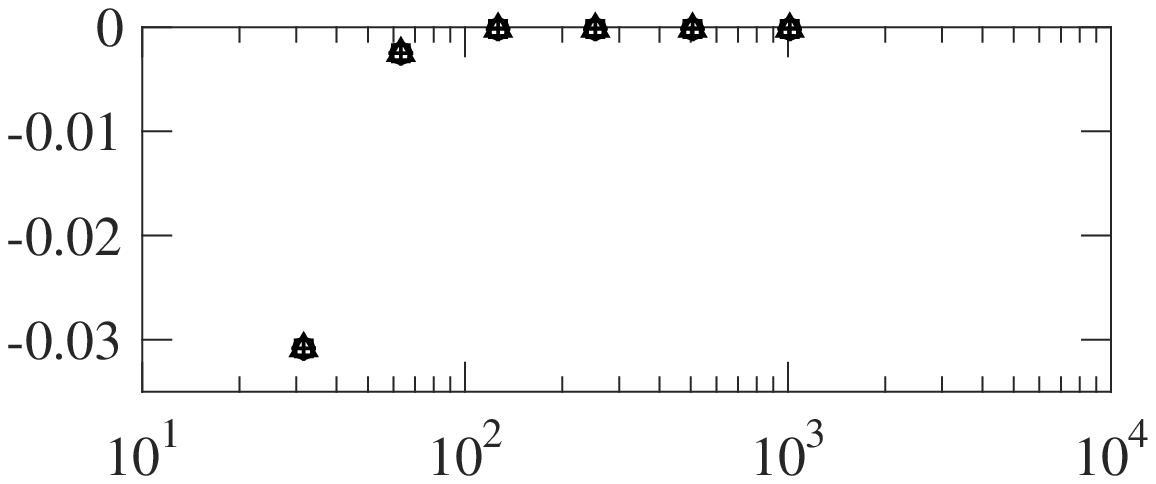}}
  \subfigure{\includegraphics[width=0.49\textwidth]{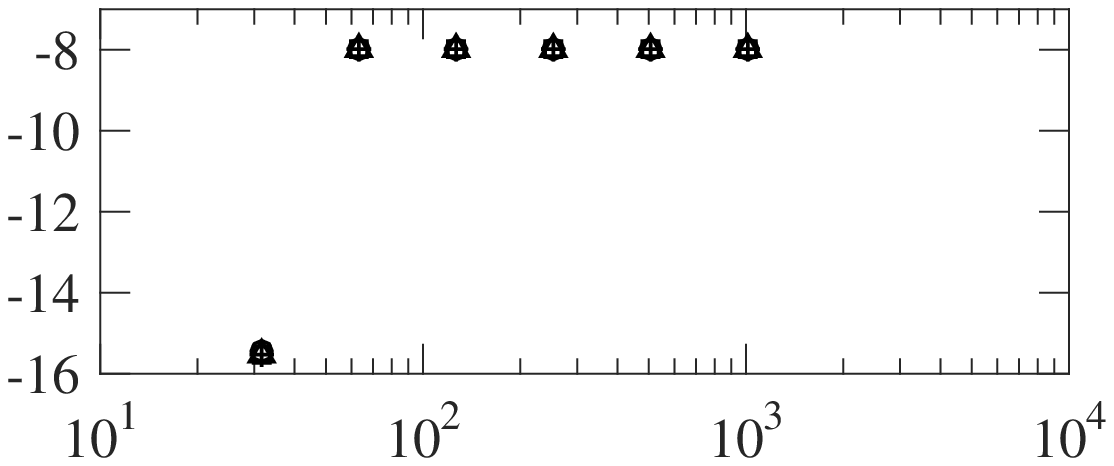}}
  \subfigure{\includegraphics[width=0.49\textwidth]{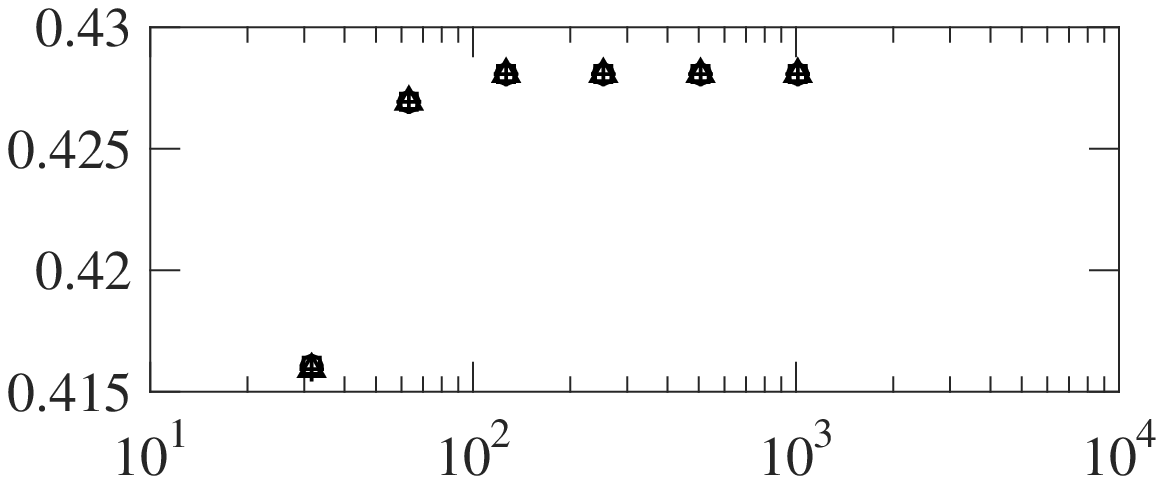}}
  \subfigure{\includegraphics[width=0.49\textwidth]{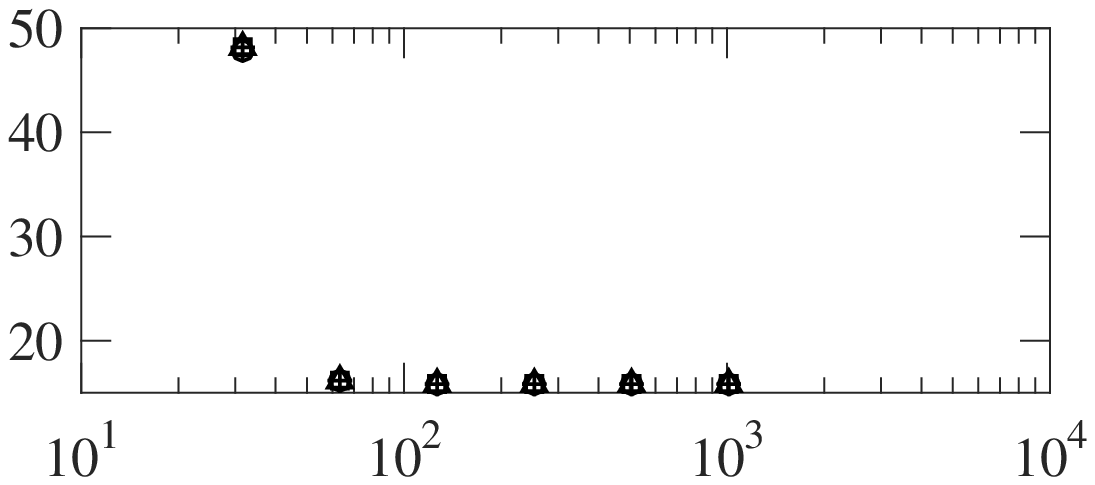}}
   \begin{picture}(0,0)
      \put(-260,0){$N$}
      \put(-85,0){$N$}
      \put(-260,73){$N$}
      \put(-85,73){$N$}
      \put(-260,145){$N$}
      \put(-85,145){$N$}
      %\put(-340,70){\rotatebox{90} {$u$}}
      %\put(-170,55){\rotatebox{90} {$u-u_a$}}
      \put(-280,175){{$\xave{u-\xave{u}}^2$}}
      \put(-280,100){{$\xave{u-\xave{u}}^3$}}
      \put(-280,25){{$\xave{u-\xave{u}}^4$}}
      \put(-130,175){{$\xave{\pd u/\pd x -\xave{\pd u/\pd x}}^2$}}
      \put(-130,100){{$\xave{\pd u/\pd x -\xave{\pd u/\pd x}}^3$}}
      \put(-130,25){{$\xave{\pd u/\pd x -\xave{\pd u/\pd x}}^4$}}
      \put(-210,195){(a)}
      \put(-38,195){(b)}
      \put(-210,117){(c)}
      \put(-38,117){(d)}
      \put(-210,45){(e)}
      \put(-38,45){(f)}
   \end{picture}
  \caption{Variation of normalized central moments of velocity and velocity-gradients with grid resolution. Computations are done using Case 6 in \rtab{cases}. Graphs (a), (c) and (e) show the second, third and fourth moments of velocity, respectively. Graphs (b), (d) and (f) show the second, third and fourth moments of velocity-gradients, respectively. Symbols represent the probability sets $\{p_0,p_1,p_3\}$:  $\{1,0,0\}$ (circle), $\{0.7,0.2,0.1\}$ (plus), $\{0.4,0.4,0.2\}$ (square), $\{0.3,0.5,0.2\}$ (triangle).}
  \label{fig:at-conv}
\end{figure}

In nonlinear problems like turbulence, one is interested not only in 
statistical moments of 
the velocity field but also in its gradients as they exhibit very strong 
but localized fluctuations, a phenomenon known as intermittency \cite{SA97}.
%To understand further the effect of asynchrony, thus,
Thus, we investigated 
the variation of central moments of velocity ($u$) and velocity gradients
($\pd u /\pd x$) with the resolution, $N$. 
An example is shown in \rfig{at-conv}.
For this problem, most of the contribution to the velocity field
comes from low wavenumbers, while most of the contribution for its 
gradients comes from high wavenumbers.
Thus, it is not surprising that asynchrony effects are more evident for velocity
gradients at low grid resolution.
Nevertheless, both synchronous and  asynchronous cases seem to converge at
the same grid resolution ($N=256$).  This is consistent with the results 
for lower-order statistics in \cite{DA2014}.

While our numerical experiments show accurate results even for high
order statistics of velocity gradients, this result is not 
expected to be general for other equations or in higher-dimensional
spaces. This is indeed an area that needs further investigation.

\section{Conclusions}

A number of natural and engineering systems are governed by PDEs
that present solutions with wide range of scales which 
can only be captured by 
%high-order numerical schemes 
high-fidelity
simulations (using high-order numerical schemes) on massive computational systems.
At extreme scales, global communications and synchronizations will likely 
become an obstacle to sustained performance for number of scientific codes.
In this work we have presented a general methodology to analyze 
and derive schemes that remove these two main obstacles by allowing 
some tunable level of asynchrony.

The concept relies on finite differences to approximate 
derivatives of general order using values of 
the function from neighboring points. Close to PEs
boundaries current computational methodologies are stalled until communications
between PEs is completed. In previous work 
we have shown that one can relax this 
forced synchronization at the mathematical level such that 
computations can proceed using values from past time levels. 
Here we generalize the concept, established conditions 
under which schemes can be obtained, classified the 
resulting schemes in terms of their properties,
and provided a general framework in which
schemes of arbitrary order can be obtained for any derivative of 
a function. These schemes are referred as asynchrony-tolerant or \ats schemes.

By analyzing in detail 
the truncation error of general finite differences 
when asynchrony 
is allowed, we described the mathematical conditions needed to 
obtain a scheme of arbitrary order under asynchronous conditions.
In particular, we showed that asynchrony errors 
can be eliminated either by extending the stencil in 
space as well as in time. These two alternatives lead
to
schemes with different properties and limitations.
However, depending on the order of the scheme and the 
type of asynchrony allowed (e.g.\ on both sides of the stencil,
uniform across the stencil, etc.) not all expansions of the 
stencil size will result in an \ats scheme.
The kind of stencils that do lead to a scheme,
has also been presented and requires identifying 
the nature of terms present in the truncation error.
The coefficients are obtained by solving a linear
system of equations.
An alternative method was also presented where 
successive terms in the truncation errors are 
eliminated by a step-by-step method. The process
ends when the desired accuracy is achieved.

The resulting schemes can be classified on the nature of
their coefficients. We presented four conditions 
for the classification: 
(i) symmetric layout of grid points,
(ii) unconstrained or uniform delay on boundary points, 
(iii) artificial delay at interior points, and 
(iv) symmetry on the coefficients. 
Each have 
different numerical and performance properties.
Actual examples of these different schemes were also 
put forth.

The truncation error was analyzed in a statistical 
framework that takes into account both the stochasticity 
of delays as well as the non-uniformity of delays in 
space. We have further shown that multi-step time-integration
methods can be used successfully to obtain solvers of 
arbitrary order using AT schemes. The general form of
the error, given by \req{aveE_scaling_gen}, shows that the average
error depends on the number of processors as well as moments of 
the distribution of delays, which in turn depend on the
characteristics of the computing system simulations are run on.

Theoretical predictions on the accuracy of schemes were compared to 
numerical experiments for
linear as well as non-linear equations. Good agreement was found across
different parameter space. The work presented here provides a strong
foundation for mathematically asynchronous computing methods for PDEs at extreme
scales. Application of this method to more complex phenomena and realistic
conditions is a part of our ongoing research.

\section{Acknowledgments}

The authors gratefully acknowledge NSF
(Grant OCI-1054966 and CCF-1439145)
for financial support. The authors also
thank NERSC and XSEDE for computer time on their systems. The authors benefited
from
discussions with Lawrence Rauchwerger, Raktim Bhattacharya and
Jacqueline H.\ Chen.

% \appendix
% \section{}
% \label{}

% References
%
% Following citation commands can be used in the body text:
% Usage of \cite is as follows:
%   \cite{key}          ==>>  [#]
%   \cite[chap. 2]{key} ==>>  [#, chap. 2]
%   \citet{key}         ==>>  Author [#]

% References with bibTeX database:

\bibliographystyle{model1-num-names}
\bibliography{main.bib}

% Authors are advised to submit their bibtex database files. They are
% requested to list a bibtex style file in the manuscript if they do
% not want to use model1-num-names.bst.

% References without bibTeX database:


% \bibitem must have the following form:
%   \bibitem{key}...
%


% \end{thebibliography}

\end{document}